\documentclass[prl,article,twocolumn]{revtex4-1}%
\usepackage{graphicx}
\usepackage{subfigure}
\usepackage{amsmath}
\usepackage{ulem}
\usepackage{color}
\setcitestyle{round}

\begin{document}
\title{Effect of Sequence-Dependent Rigidity on Plectoneme Localization in
dsDNA}
\author{Shlomi Medalion}
\author{Yitzhak Rabin}
\affiliation{Department of Physics and Institute of Nanotechnology and Advanced
Materials,
Bar-Ilan University, Ramat-Gan 52900, Israel}
\date{\today}

\begin{abstract}
We use Monte-Carlo simulations to study the effect of variable rigidity on plectoneme formation and 
localization in supercoiled dsDNA. We show that the
presence of soft sequences increases the
number of plectoneme branches and that the edges of the branches tend to be
localized at these sequences. We propose an experimental approach to test our
results in vitro, and discuss the possible role played by plectoneme localization in the search process
of transcription factors for their targets (promoter regions) on the
bacterial genome.
\end{abstract}

\maketitle

The conformational properties of double-stranded (ds) DNA molecules are usually modeled by 
treating these biopolymers as semi-flexible
chains with uniform rigidity that can be represented by a single persistence length 
(rigidity is proportional to persistence length),  $l_p\simeq50nm$ in physiological conditions 
\cite{hagerman1988flexibility,taylor1990application}. This is 
justified when one is interested in  large-scale properties of dsDNA for
which one can replace the sequence-dependent distribution of the elastic rigidity 
by its average over the chain. When one is interested in small and intermediate-scale phenomena
one has to consider the full sequence-dependence of the rigidity. While some studies suggested 
that the rigidity of bare dsDNA varies across a limited range $40nm<l_p<75nm$
\cite{leger1998reca, geggier2010sequence}, 
experiments on cyclization of short dsDNA fragments ($\approx 100 bp$) reported
much higher cyclization
ratios than expected \cite{cloutier2004spontaneous,cloutier2005dna}. This led to the proposal
of DNA kinks -- pointlike highly flexible domains \cite{wiggins2005exact} --
perhaps due to formation of ``DNA bubbles'' 
\cite{yan2005statistics,yuan2006spontaneous}. Such
sequence-dependent rigidity is a property of bare dsDNA and it has  
been suggested that the effect can be utilized for the design of  promoter sequences in order to
control the DNA binding affinity of transcription factors 
that are sensitive to DNA bendability \cite{levo2014pursuit}. It may also arise
as the consequence of the binding of proteins to specific DNA sequences; indeed,
in vivo
DNA is partially covered by proteins that affect its flexibility
\cite{williams2010biophysics, rappaport2008model, amit2003increased,
leger1998reca, egelman1986structure} and/or its local curvature
\cite{jones1999protein, stavans2006dna}. For example, RecA bacterial
proteins polymerize along DNA to give an effective persistence length of hundreds of
nm to the RecA-dsDNA complex
\cite{egelman1986structure, leger1998reca}. Other positively-charged proteins
(e.g., HMGB) and polyamines, increase DNA's flexibility
significantly \cite{williams2010biophysics, podesta2005positively,
hegde2002papillomavirus}.  Thus, the variability of DNA rigidity may be even higher
in vivo than that of bare DNA in vitro. 

\begin{figure}[tb]
\center{\includegraphics[width=0.45\textwidth]{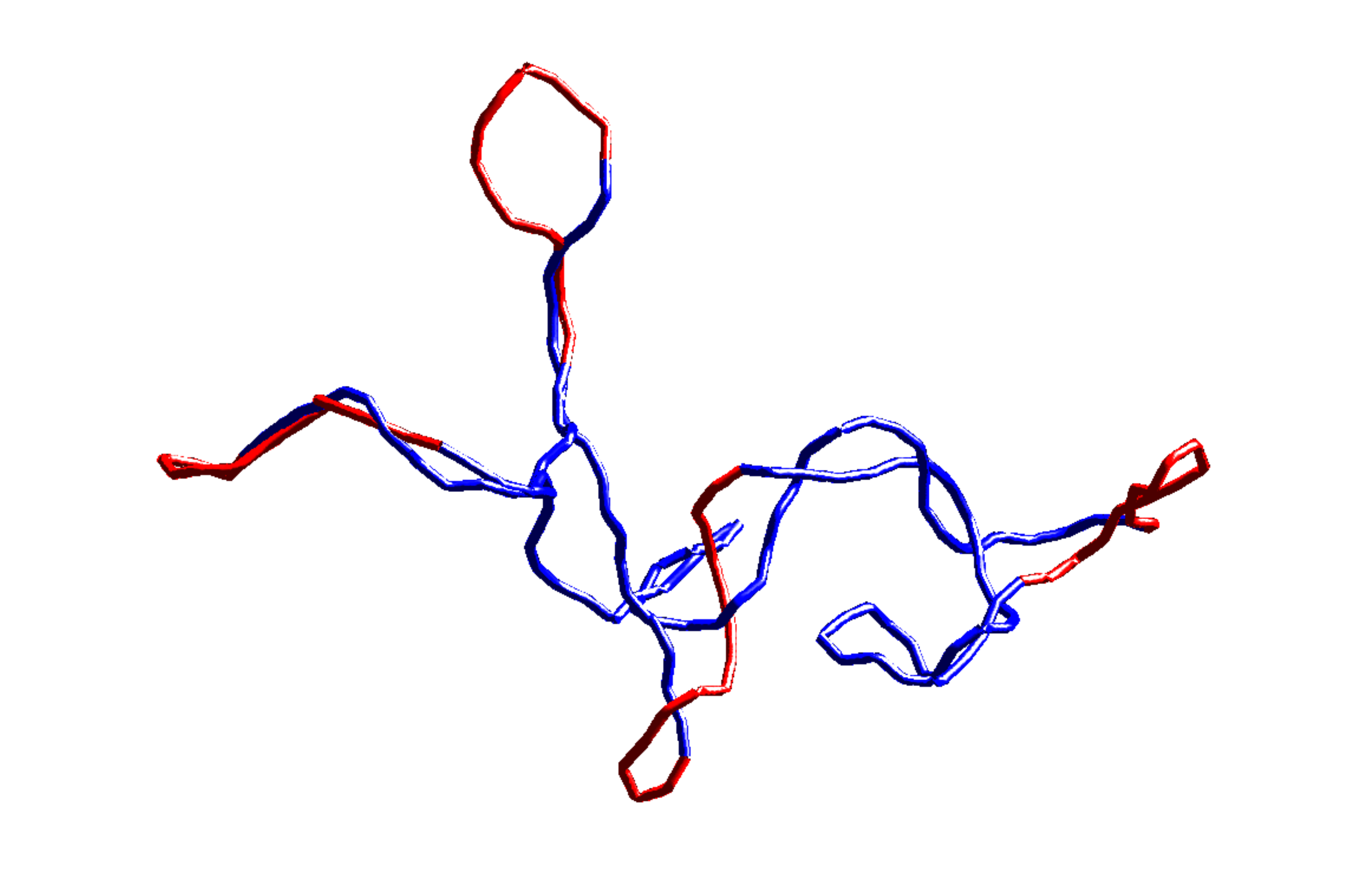}}
\caption{(color online) Representative conformation of a supercoiled chain of
length of $5400bp$ with a pair of low and high rigidity values $(l_p^{(l)}, l_p^{(h)})
= (55nm, 75nm)$. Here $l_p^{(l)}$ occupies $1/3$ of the chain and is  divided into $4$
equally-spaced domains of identical size, $l_p^{(h)}$ occupies the remaining
$2/3$ of the chain, and the superhelical density is $\sigma=-0.05$. The red
segments represent the soft domains, while the blue ones correspond to the
stiffer ones. The end loops of the plectonemes tend to be localized at the
soft domains.}
\label{fig:Fig1}
\end{figure}

In this work we study the interplay between local rigidity and
\textit{plectoneme localization} in supercoiled dsDNA.
When circular DNA is subjected to sufficiently large torsional stress, the minimization
of the free energy yields strongly-writhed conformations known as
plectonemes (see e.g. Fig. \ref{fig:Fig1}). We show that the number of
plectonemic branches and the locations of the edges (end loops of the
branches in Fig. \ref{fig:Fig1}) are affected by non-uniform DNA
rigidity. This applies not only to circular chains
(e.g., bacterial and mitochondrial DNA, and DNA plasmids), but to
topologically-constrained linear dsDNA molecules as well, such as eukaryotic
chromosomes that are attached to the nuclear lamina.

We performed Metropolis Monte-Carlo (MC) simulations of topology-conserving
worm-like rod (WLR) model that accounts for bending and twist elasticity
of dsDNA.
A circular dsDNA molecule was modeled as a closed chain of
$N$ segments of length $30bp\simeq10nm$ each. The
twist persistence length of the chain was taken to be $l_{tw}=74nm$ (corresponding to
$\tilde{l}_{tw}=7.4$ in units of segment length). The effective diameter of
dsDNA was assumed to be $5nm$, taking into account the screened electrostatic
repulsion at physiological conditions
\cite{rybenkov1997effect,vologodskii2009simulation}. The degree of supercoiling
of DNA was characterized by the \textit{superhelix density} $\sigma$ that is proportional to
the amount of torsional stress per unit length of the molecule
\cite{vologodskii1994conformational, vologodskii2009simulation}.
Bacterial cells have an enzymatic mechanism that keeps the superhelix density
of the genome at an almost constant value, typically in the range  $-0.03<\sigma<-0.09$
that corresponds to somewhat unwound DNA
\cite{bauer1978structure,vologodskii1994conformational} and,
unless stated otherwise, we assumed that $\sigma=-0.06$.
For more details about the energy form used for accepting/rejecting the
Metropolis MC steps see Sec. S1 in the SI.

In  the simulations we used the pivot (or crankshaft) moves  described in
detail in ref. \cite{vologodskii1992conformational}. The angle of rotation
around the pivot was tuned in order to achieve the
desired
acceptance rate of $50\%$. To make sure that the simulation was not stuck in a
specific plectonemic conformation, we used the \textit{linking number inversion}
method described in ref. \cite{medalion2014effect}.
We ran our simulations for $5400bp$ long chains with segments of  two different
rigidities distributed along the chain such that
most of the chain had a higher rigidity ($l_p^{(h)}$), and a smaller part of the 
chain had a lower rigidity ($l_p^{(l)}$). 
For each chain conformation taken into account, we determined the number and
the locations of plectoneme edges (end loops). The algorithm used for this analysis
is described in the SI.
The chains in the simulations were unknotted and we made sure that this
topology was fixed during the simulation by calculating the Alexander
polynomials and Vassiliev invariants after each move, and rejecting
topology-changing moves.

\begin{figure}[tb]
\center{\includegraphics[width=0.5\textwidth]{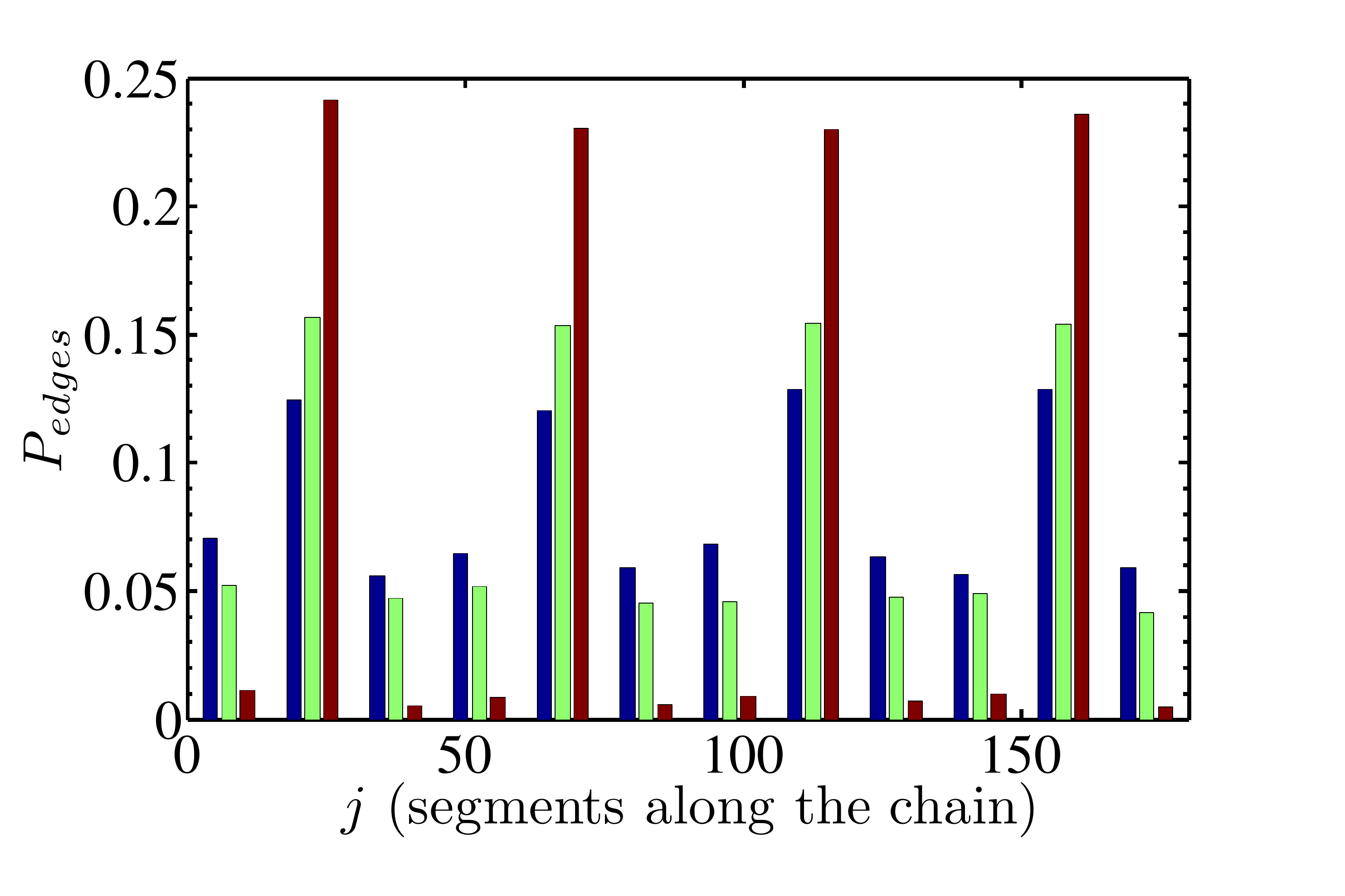}}
\caption{(color online) Histogram of the locations of the centers of
plectoneme edges (end loops), $P_{edges}$, for a circular chain
with $\sigma=-0.06$ and $4$ soft domains located around the $23$rd, $68$th,
$113$rd and $158$th segments. The rigidities correspond to the
pairs $(l_p^{(l)}, l_p^{(h)}) = (40nm, 55nm)$ (blue), $(55nm, 75nm)$ (green),
and $(50nm, 200nm)$ (red).}
\label{fig:Fig2}
\end{figure}

We first considered the effect of non-uniform rigidity
distribution on plectoneme localization in the case where both the stiff and
the soft domains are relatively large (as for example in the RecA-DNA
complexes).
The soft domain was chosen to occupy $1/3$ of the chain and consisted of
several domains of identical size, equally-spaced along the contour. 
In order to model sequence-dependent  
rigidity of bare dsDNA we took $(l_p^{(l)}, l_p^{(h)}) = (40nm, 55nm)$
(as in  ref. \cite{geggier2010sequence}) and $(55nm,
75nm)$ (as in  ref. \cite{leger1998reca}). For modeling rigidity induced 
by protein polymerization along DNA we used $(l_p^{(l)}, l_p^{(h)}) = (50nm,
200nm)$ (even though polymerization of proteins such as RecA yields rigidities
in the range of $600nm-800nm$ \cite{egelman1986structure,leger1998reca}, the
effect of $l_p^{(h)}\gg l_p^{(l)}$ on the localization of plectoneme edges is
already clear for $l_p^{(h)}=200nm$).
In Fig. \ref{fig:Fig2} we plotted a histogram of the locations of the centers of
plectoneme edges for a chain with $4$ soft domains.
Even for the pair with the smallest difference between rigidities  $(40nm,
55nm)$, there is noticeable
localization of the edges to the softer segments of the chain. As the difference between rigidities increases, the
localization becomes much more pronounced. For $(l_p^{(l)}, l_p^{(h)}) =
(50nm, 200nm)$, the edges are almost exclusively located in the soft
domains. While high superhelix density is a necessary condition
for the formation of plectonemes \cite{vologodskii1994conformational}, 
we found that increasing $\sigma$ beyond the plectoneme formation threshold does
not have a noticeable effect on the localization of the edges of these branches
(see Fig. S5 in the SI).

For chains of uniform rigidity the number of plectonemic branches was
shown to increase with $\sigma$ \cite{vologodskii1992conformational}.
In order to understand how the dependence of the number of chain branches on
$\sigma$ is affected by non-uniform rigidity,
we compared chains with $4$ soft domains with uniform chains of a single
persistence length (weighted average of the pair of the corresponding 
persistence lengths). We observed that for
the $(50nm, 200nm)$ pair, the presence of soft
domains leads to two-fold enhancement compared to homogeneous chains with
an average value of $l_p=150nm$ (the effect is much weaker both for chains with lower rigidities
and those with a smaller difference between higher and lower rigidities- see Fig.
S3 in the SI).
The origin of this effect is that a chain with $l_p=150nm$ has to pay a
large bending
energy penalty in order to create a tight loop/edge and, as a consequence, the number of
edges is small. The presence of soft domains with $l_p=50nm$ (even when
the stiffness of the rest of the chain is increased), allows the edges to be
localized in these domains, thus
decreasing the bending energy penalty, and the number of branches increases significantly.
The number of plectoneme branches is also affected by the number of soft
domains. We calculated the average number of
branches as a function of the number of soft domains for
$(40nm, 55nm)$, $(55nm, 75nm)$ and $(50nm, 200nm)$, 
where the softer part of the chain was divided into $2-6$ domains
of equal length, uniformly distributed along the chain (Fig. S4 in the SI).
As expected, only a minor effect is observed in the first two cases where the
difference between rigidities is small but in the case of large rigidity
contrast, the number of edges increases by nearly $50\%$ as the number of soft
domains increases from $2$ to $6$.

\begin{figure}[tb]
\center{\includegraphics[width=0.5\textwidth]{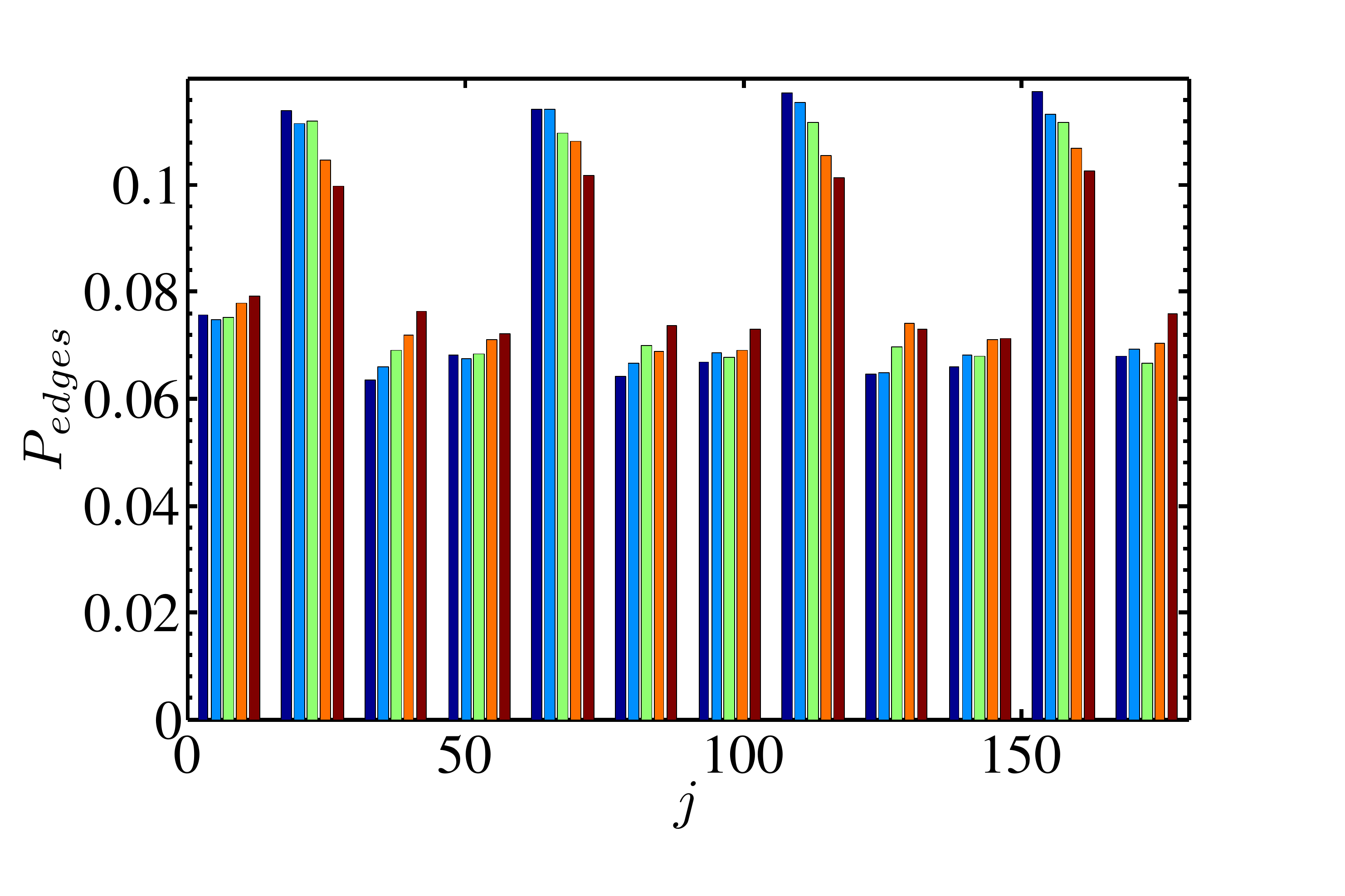}}
\caption{(color online) Histogram of the locations of plectonemic edges,
$P_{edges}$, for chains with $4$ kinks located at the $23$rd, $68$th,
$113$rd and $158$th dimers, with $\sigma=-0.04$ (dark blue), $\sigma=-0.05$
(light blue), $\sigma=-0.06$ (light green), $\sigma=-0.07$ (orange) and
$\sigma=-0.08$ (red).}  
\label{fig:Fig7}
\end{figure}

Next, we proceeded to characterize the effect of DNA kinks (soft local defects) 
on the localization and the number of plectoneme edges. We simulated $5400bp$-long chains
with $l_p=50nm$, and with $2-6$ kinks at which the persistence length dropped
to a value of $4nm$ (about twice the rigidity of ssDNA,
\cite{murphy2004probing, yuan2006spontaneous}).
As can be seen in Fig. \ref{fig:Fig7} for the $4$-kink case (and 
for $2-6$ kinks in Fig. S6 in the SI), plectoneme edges tend to form
at the locations of kinks. Another effect observed in Fig. \ref{fig:Fig7}
is that in the $4$-kink case the fraction of edges that contain kinks decreases monotonicaly
with $|\sigma |$, presumably because of the number of branches increases monotonically with
$|\sigma |$. In the $6$-kink case the decrease in the fraction of kink-containing edges is 
observed only for $|\sigma| = 0.07-0.08$ (Fig. S9 in the SI). This concurs with
the observation (Fig. S8 in the SI) that while in
the $6$-kink case the number of edges is only slightly greater than the number
of kinks in the chain even for the highest value of $|\sigma| $, 
in the $4$-kinks case the number of edges exceeds
the number of kinks already for moderate $|\sigma |$s.

In Table \ref{table1} and in Figs. S6-S7 in the SI, we show that as
the number of kinks, $n_{kink}$, increases, the number of
branches, $n_{branch}$, also grows, but this increment is accompanied by
an even more rapid increase in  the fraction of 
edges that include a kink, $P_{kink}^{(tot)}$, and in the total number of edges containing a kink, 
$n_{\cap}$. Even though plectonemes have branches even in the absence
of kinks, when supercoiled DNA contains kinks, the edges of the branches
are localized preferentially at the kinks; this reduces the bending energy
penalty for creating edges and promotes the formation of new branches. 

\begin{table}[tb]
\centering % used for centering table 
\begin{tabular}{c c c c} % centered columns (4 columns) 
 \hline\hline %inserts double horizontal lines 
$n_{kink}$ & \quad $\langle n_{branch}\rangle$  & \quad $P_{kink}^{(tot)}$ 
\quad & \quad $\langle n_{\cap}\rangle$ \qquad  \\ [0.5ex]
\hline % inserts single horizontal line 
$2$\quad & $5.99$ & \quad$0.332$\qquad & \quad $1.98$ \qquad  \\
$3$\quad & $6.28$ & \quad$0.399$\qquad & \quad $2.50$ \qquad   \\
$4$\quad & $6.41$ & \quad$0.441$\qquad & \quad $2.63$ \qquad   \\
$5$\quad & $6.62$ & \quad$0.510$\qquad & \quad $3.37$ \qquad  \\
$6$\quad & $6.76$ & \quad$0.544$\qquad & \quad $3.67$ \qquad   \\
\hline %inserts single line
\end{tabular}
\caption{The number of branches, $\langle
n_{branch}\rangle$, the total
probability of an edge to contain a kink, $P_{kink}^{(tot)}$, and the total
number of edges containing a kink, $\langle n_{\cap}\rangle = \langle
n_{branch}\rangle\cdot P_{kink}^{(tot)}$
as a function of the number of kinks in the chain, $n_{kink}$,
for $5400bp$ long chains ($\sigma=-0.06$).}
\label{table1}
\end{table} 

%When the torsional stress is increased at a fixed number of kinks,
%initially the number of plectoneme branches is smaller than the number of kinks and most of 
%the edges will contain kinks. Upon further increase of $|\sigma |$, 
%the number of plectoneme branches (and therefore of edges) will eventually exceed
%the number of kinks and the fraction of edges containing a kink will decrease.

\begin{figure}[tbh]
\centering
\subfigure[] { \label{fig:fig4dist00}
\includegraphics[scale=0.37]{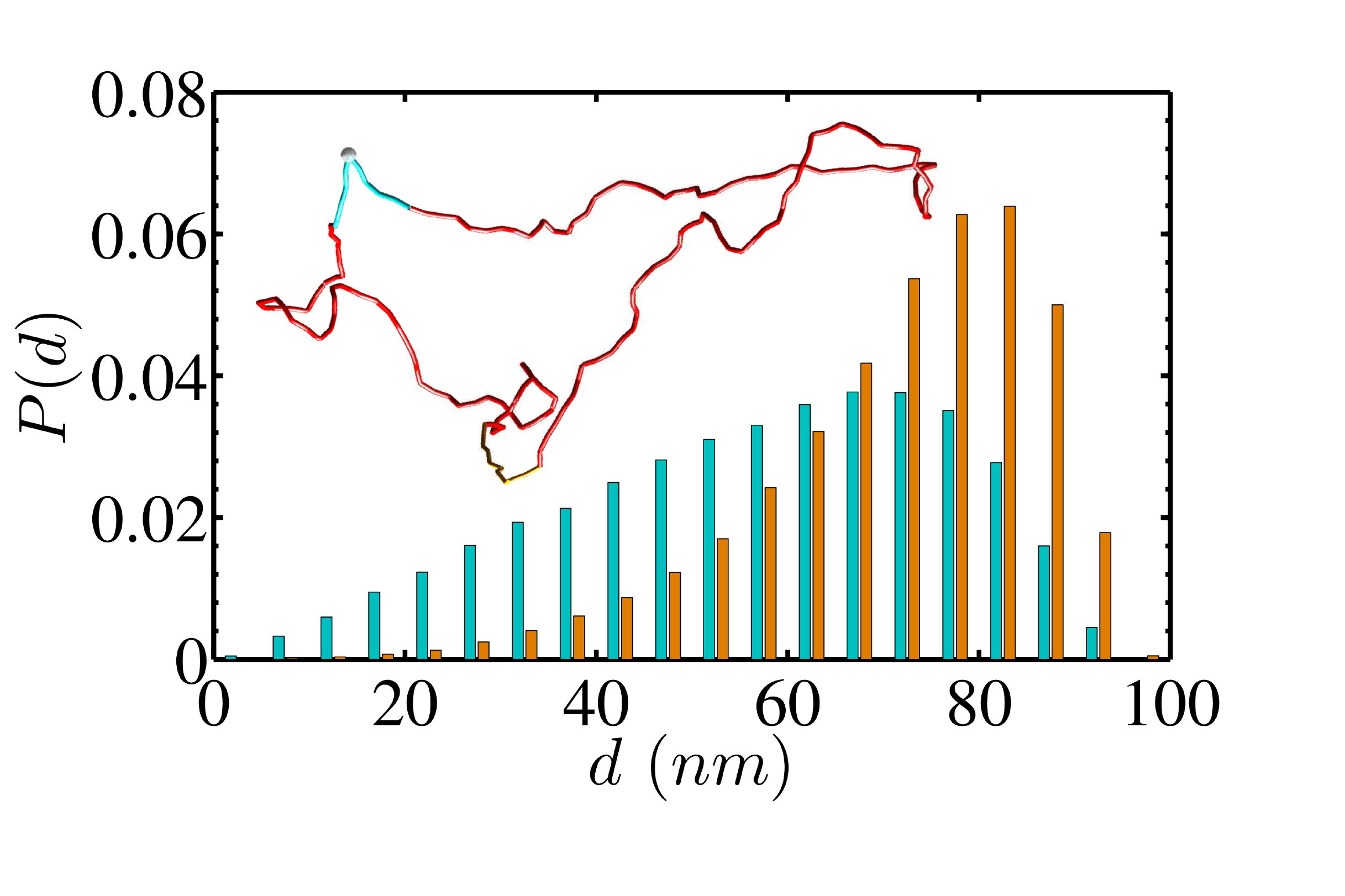}}
%\newline%
\centering
\subfigure[]{\label{fig:fig4dist06}
\includegraphics[scale=0.37]{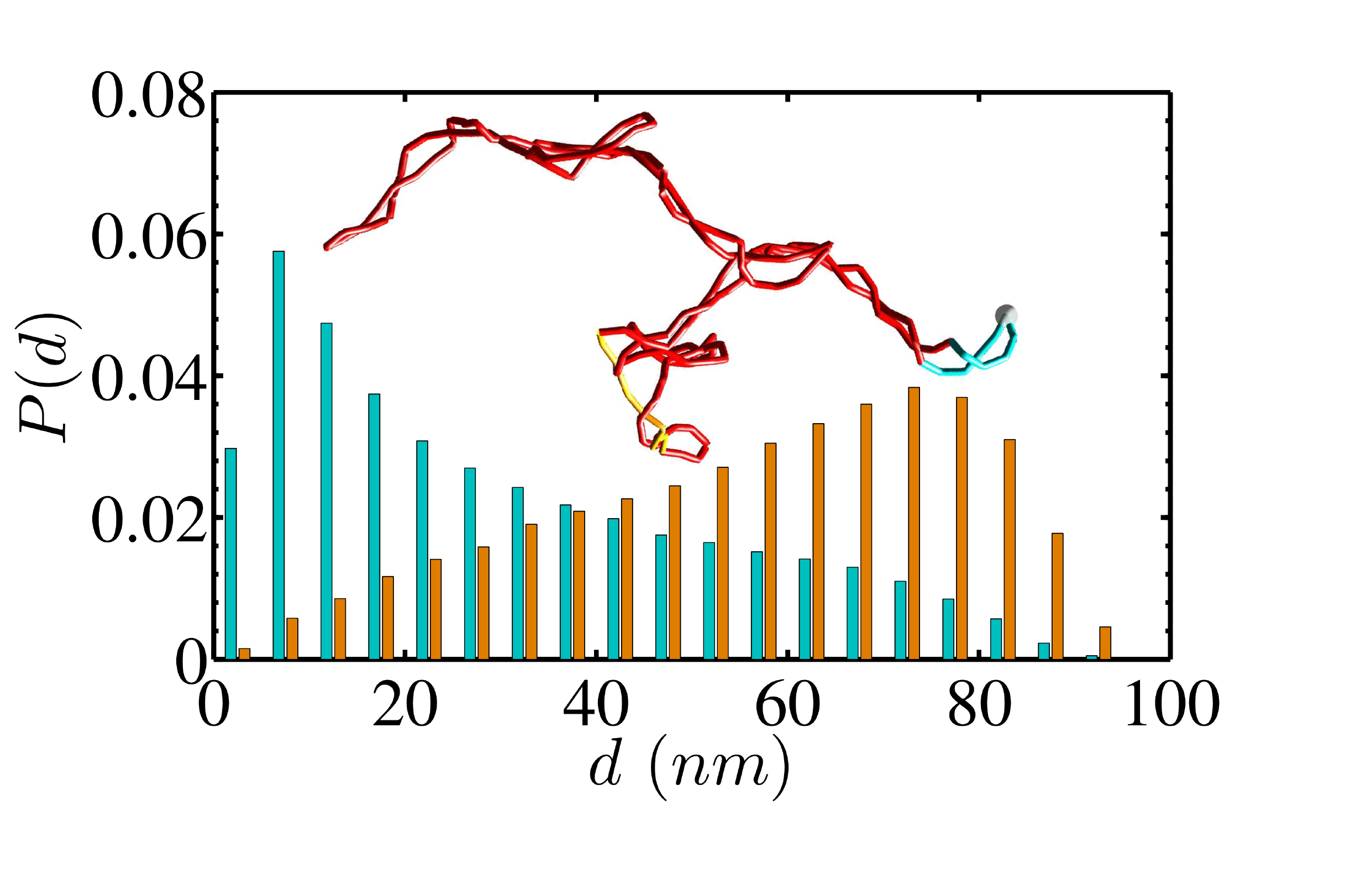}}
\caption{(Color online) Histograms of distances of the end points of a
$100nm$ long segment in a $5400bp$ plasmid containing a kink (light blue) and
not containing a kink (orange), for (a) torsionally relaxed ($\sigma=0$)
plasmids and (b) torsionally stressed ($\sigma=-0.06$) plasmids. 
Representative DNA conformations are shown in the corresponding figures (red), with
a sequence (light blue) containing a kink (sphere).
A sequence of the same length but without a kink is shown for comparison
(orange.)}
\label{fig:Fig4}
\end{figure}

As expected from experience with linear worm-like chains, the presence of a kink
affects the local fluctuations of the chain in which it is imbedded. This effect
is clearly observed in Fig \ref{fig:fig4dist00} where  we compare the
distribution
of the end-to-end distance $d$ of a short DNA sequence (in a long circular
chain) of uniform rigidity, with that of a DNA sequence 
of the same length and rigidity but containing a kink at its center. Since the
length of the sequence is only twice the
persistence length of dsDNA, both distributions are asymmetric, with a peak closer to the fully stretched ($100nm$) state.
The introduction of the kink shifts the maximum of the distribution from $80nm$ to $70nm$ and results in dramatic enhancement
of the ends contact probability ($d\leq 10nm$), an effect that has been invoked
to explain the anomalously high cyclization probability 
of short dsDNA molecules\cite{cloutier2004spontaneous,cloutier2005dna,wiggins2005exact}. The difference between the distributions
becomes much more pronounced in the presence of supercoiling. Since the end
loops of plectonemes tend to form
around the kinks and the size of the end loop is of order $l_p$, we expect
the ends of the $100nm$  segment to be 
preferentially localized in the stem of the plectoneme and the most probable end-to-end separation will be below $10nm$. This expectation is
fully confirmed in Fig. \ref{fig:fig4dist06} in which we presented the corresponding end-to-end distance distributions 
of supercoiled DNA for $\sigma=-0.06$  (the histograms for $\sigma$ in the range
between $0$ and $-0.07$ are shown in Fig. S10 in the SI). While
the distribution for the no-kink case is qualitatively similar to that for $\sigma=0$ (Fig \ref{fig:fig4dist00}), the peak of the distribution 
of a  kink-containing segment is shifted towards contact. In principle, this prediction can be tested by FRET experiments by attaching donor and acceptor molecules to the ends of a short (of order $l_p$) sequence in a long circular dsDNA molecule, provided that one is able to design a sequence that has a high propensity for forming a kink \cite{mcnamara1990sequence,kim1993co,peyrard2009experimental,cuesta2009adding}. In the absence of supercoiling,
the donor and acceptor will be separated by tens of nanometers and there will be no FRET signal. As torsional stress is introduced into the chain, e.g., by raising the concentration of intercalators in the solution \cite{bauer1968interaction,steck1984dna}, plectonemes will form with end loops localized at the kinks
and the distance between the donor and acceptor will approach the $6nm$ limit at which energy transfer will take place and a FRET signal will be observed.

This localization effect may also play a significant role in the search
process executed by transcription factors (TFs) that attach to specific
sequences in promoter regions of DNA. Since bacterial DNA is supercoiled, DNA
segments that reside in the stems of the plectonemes are pressed against each other and
one expects binding of proteins to these segments to be suppresed. The looped ends of plectonemes
are more accessible and, combined with the fact that many bacterial DNA-associated proteins have a 
higher binding affinity to bent DNA \cite{kim1989bending, stavans2006dna,
medalion2012binding}, this suggests that TFs may have a higher affinity for end loops of plectonemes.
Whether nature utilizes such effects to facilitate the search of TFs for their DNA binding sites
depends on whether promoter sites tend to be localized at the edges of plectonemes.
Indeed, many bacterial promoters contain TATA-box sequences that are known
for their lower rigidity and higher tendency to create DNA bubbles (kinks)
\cite{mcnamara1990sequence, peyrard2009experimental,cuesta2009adding,kim1993co}
and, according to our results, such sequences will tend to nucleate plectonemes and to
be positioned at their edges. Finally, we would like to mention that although we are
not aware of direct experimental proof of our proposed mechanism of plectoneme localization 
to low-rigidity DNA sequences, there has been a recent experimental study of the effects of torsional 
stress on stretched linear dsDNA in which repeated
hopping of plectonemes between specific locations along DNA was reported.
\cite{van2012dynamics}.

\section{Acknowledgments}
\label{sec:acknowledgments}

Results obtained in this paper were computed on the biomed virtual organization
of the European Grid Infrastructure (http://www.egi.eu). We thank the European
Grid Infrastructure and supporting National Grid Initiatives (listed here:
http://lsgc.org/en/Biomed:home \#Supporting\_National\_Grid\_Initiatives) for
providing the technical support, computing and storage facilities. This work was
supported by the I-CORE Program of the Planning and Budgeting committee and the
Israel Science Foundation,
and by the US-Israel Binational Science Foundation.

%\bibliographystyle{plain}
%\bibliography{bibliography.bib}

\bibliographystyle{apsrev4-1}
\bibliography{bibliography}

\newpage

\renewcommand{\thepage}{S\arabic{page}}  
\renewcommand{\thesection}{S\arabic{section}}   
\renewcommand{\thetable}{S\arabic{table}}   
\renewcommand{\thefigure}{S\arabic{figure}}
\renewcommand{\theequation}{S\arabic{equation}}
\setcounter{figure}{0}

\onecolumngrid

\section{Supporting Information}

\subsection{The Energy Form for the MC Process}

In a discrete worm-like rod (WLR) model that accounts for both the bending
diversity and twisting elasticity of the DNA, the energy of a chain (in units
of $k_BT$) with $N$ segments is given by:
\begin{equation}
\frac{E_{WLR}}{k_{B}T}=\sum_{n=1}^{N}\left[\frac{\tilde{l}_{p,n}}{2}
(\tilde{\kappa}_{n})^{2} +
\frac{\tilde{l}_{tw}}{2}({\delta\tilde{\omega}_{tw,n})^{2}}\right],
\label{eq:WLR_Energy_Discrete}%
\end{equation}
where $\tilde{l}_{p,n}$ is the dimensionless bending persistence length
corresponding to the ($n$, $n+1$) dimer of segments,  $\tilde{l}_{tw} $  is the
dimensionless
twist persistence length (both measured in units of the segment's length $\Delta
s$), and $\tilde{\kappa}_{n}=2[1-cos(\theta_{n,n+1})]$ is the dimensionless
curvature between the $n$th and the $(n+1)$th segments.
$\delta\tilde{\omega}_{tw,n} =
\tilde{\omega}_{tw,n} - \tilde{\omega}^{(0)}_{tw}$ defines the difference
between
the $n$th twist angle, $\tilde{\omega}_{tw,n}$, and its spontaneous value,
$\tilde{\omega}^{(0)}_{tw}$. The twist angle is the sum of the first and the
third rotation angles ($\tilde{\omega}_{tw,n} = \varphi + \psi$) of the Euler
transformation that rotates the $n$th segment to the $(n+1)$th segment, while
the bending angle, $\theta_{n,n+1}$, is the second Euler rotation angle.

Topologically, supercoiled dsDNA could be described as a closed chain made of
two strands infinitesimally closed to each other. This kind of a chain obeys two
topological constraints, i.e., the closure of each of its strands. Another
description for these topological constraints is given in terms of the closure
of the center-line of the chain, and the closure of the cross-sectional plane.
These two constraints are fully accounted by the Fuller-White relation
\cite{fuller1971writhing,white1969self,medalion2010coupling}:
\begin{equation}
{Lk}={Wr}+{Tw}. 
\label{eq:Fuller}%
\end{equation}
where the writhe $Wr$ is a measure of the deviation of the center-line
from planarity, and the twist $Tw$ is proportional to the the sum
(in the discrete description) over all the twist angles along the chain. The
continuum expressions for $Tw$ and $Wr$ are given e.g., in
\cite{medalion2010coupling}. In their discrete version they take the forms:
\begin{equation}
 Tw = \frac{1}{2\pi}\sum_n \tilde{\omega}_{tw,n}, 
\label{eq:discreteTw} 
\end{equation}
and:
\begin{equation}
 Wr = \frac{1}{4\pi}\sum_{n,m} W_{n,m}, 
\label{eq:discreteWr1} 
\end{equation}
where $W_{n,m}$ is the discrete pair-of-segments contribution to the $Wr$,
calculated e.g., in \cite{klenin2000computation} ($Wr$  depends
only on the coordinates of the center-line of the chain).

For a circular chain, $Lk$ is a topological
invariant (an integer) that does not change as long as the topology of the chain is
maintained. A supercoiled dsDNA is overwound or unwound depending whether
its $Lk$ is higher or lower than the preferred value of overall twist,
$Tw^{(0)}=1/2\pi\sum_n\tilde{\omega}_{tw}^{(0)}=N\tilde{\omega}_{tw}^{(0)}
/2\pi$. When the
value of this difference, $\Delta Lk \equiv Lk-Tw^{(0)}$ differs from zero, the
chain becomes torsionally stressed and attempts to minimize its total free energy by
transforming some of its twist
energy into bending energy and, when a critical value of torsional
stress is reached, plectonemes appear in the chain.

Since the twist angles of different segments are independent, one may average the twist along the chain by calculating
$\langle{\delta\tilde{\omega}_{tw,n}}\rangle \simeq 2\pi Tw/N = 2\pi
(Lk-Wr)/N$, where $Lk$ is a pre-determined topological constant,
and the $Wr$ is calculated using the center-line coordinates of the chain. The
resulting twist energy, using $\Delta{Tw}\equiv{Tw}-{Tw}^{(0)} = \Delta{Lk} -
Wr$, is then given by:
\begin{equation}
E_T \simeq \frac{2\pi^2\tilde l_{tw}}{N}\Delta Tw^2 = \frac{2\pi^2\tilde
l_{tw}}{N}(\Delta Lk - Wr)^2.
\label{eq:twEnergy}
\end{equation}
Thus, the twist energy in our simulations is obtained
by using only the coordinates of the center-line of the chain for calculating
$Wr$, and plugging this value of $Wr$ (and the constant $\Delta Lk$) into Eq. \ref{eq:twEnergy} to
calculate $E_T$.

\subsection{Methods for Identifying Plectoneme Edge Location}

The number and locations of plectoneme edges (end loops) were determined
by two methods. In the first analysis we used the
method of calculation of local writhe, $Wr^{(loc)}$ described in ref
\cite{vologodskii1992conformational}. Although $Wr$ of a chain is
strictly defined only for closed chains, one may calculate the Gauss double
integral along shorter linear segments of length $l$ along the chain to obtain
the local writhe associated with the $j$th monomer, $Wr^{(loc)}(j)$.  As shown
in Fig. S1, the function $Wr^{(loc)}(j)$ has maxima
at end loops  (edges) of the plectoneme, and when the value of such a 
maximum exceeds a threshold, $Wr^{(threshold)}$ (the horizontal black dashed
line in the figure), this monomer is marked as an edge of the plectoneme (a
sphere on the chain in the figure). 
The height of the peaks of $Wr^{(loc)}$ depends both on the length $l$ of the
segments for which we calculate the local writhe and on the superhelix
density $\sigma$, since higher $|\sigma|$s yield tighter loops resulting
in higher local writhe values. 
Since we want the same threshold value, $Wr^{(threshold)}$, to distinguish
between edges to other segments for different $\sigma$s, $l$ must compensate
for the $\sigma$-dependence of the heights of the peaks.
While in ref \cite{vologodskii1992conformational} the authors flattened the
conformations to nearly planar ones before calculating the number of edges,
we wanted to analyze the three dimensional structures of the plectonemes, and
hence, we used a different function for $l(\sigma)$ and a different
value of $Wr^{(threshold)}$. In our simulations the choice $l=18l_p/(1+80|\sigma|)$ yielded
about the same average heights for the peaks when varying $\sigma$, and
$Wr^{(threshold)} =0.75$ was found to be an appropriate value to identify the
edges from other segments.

\begin{figure}[tb]
\center{\includegraphics[width=0.6\textwidth]{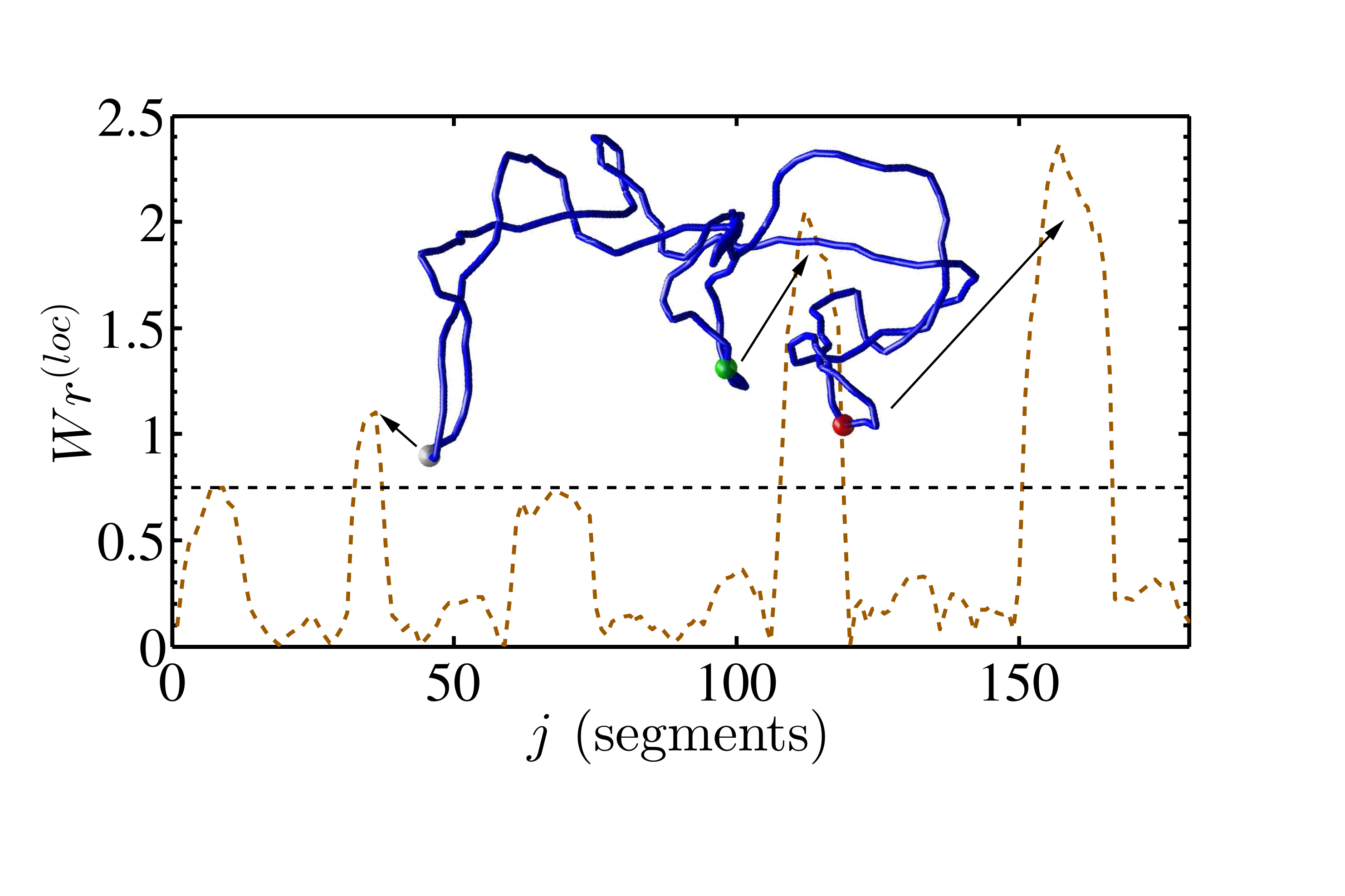}}
\caption{(Color online) $Wr^{(loc)}$ along the chain, for the specific
realization plotted in the figure. The spheres at the edges 
correspond to the locations of the peaks of $Wr^{(loc)}$.}%
\label{fig:FigS1}%
\end{figure}

For our second analysis we started from the
$j$th segment, and measured the distance $D^j(1)$ between the $(j+1)$th and
$(j-1)$th segments, the distance $D^j(2)$ between $(j+2)$th and $(j-2)$th
segments, and in general, the distance $D^j(m)$ between $(j+m)$th and $(j-m)$th
segments. Plotting $D$ for different $j$s along the chain for a specific
realization one gets curves such as in Fig. S2. If $j$ is a segment somewhere on
the stem of the plectoneme, $D$ will increase almost linearly with $m$ as
demonstrated, e.g., by the violet curve in Fig. S2, whereas $j$
corresponding to a segment in the center of an end loop will behave differently,
i.e., it will slightly increase and then will oscillate with a minimal value
very close to the (excluded) diameter of the chain,
$d$. Running along specific plectonemic conformation ($j=1,2...$), we identified
the segments that yield  local minimum of the sum of distances
$\sum_{m=1}^{m_{cut}}D^j(m)$ (for $m_{cut}=12$ and accepted only minima under a
threshold value of $20$ in units of $\Delta s$) as the segments in the center of
a loop. An advantage of this analysis is that by examining the shape of
$D^j(m)$ for the center points of an edge, one may determine the lengths of the
branches as well, since when the two chain parts depart from each other (for
some $m$), there is an abrupt change in the slope of $D^j(m)$.
However, for our purposes the two methods of analysis agreed in more than $95\%$ of the
cases.

\begin{figure}[tb]
\center{\includegraphics[width=0.5\textwidth]{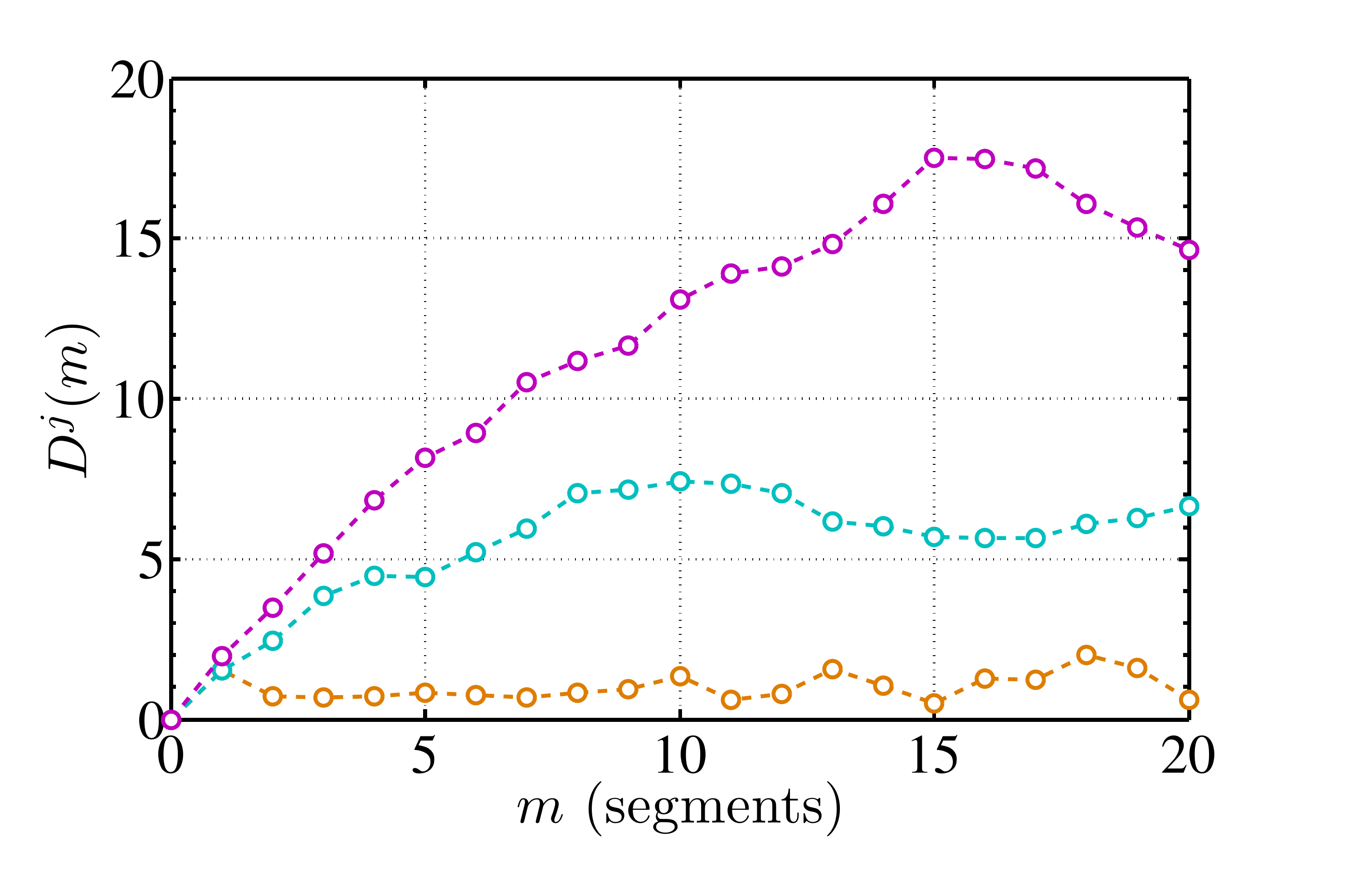}}
\caption{(Color online) Typical realization of $D^j(m)$ starting from
different $j$s along the chain: center of an edge (orange), middle of a
branch body (violet), and a point somewhere in between (light blue).}%
\label{fig:FigS2}%
\end{figure}

\begin{figure}[tb]
\center{\includegraphics[width=0.5\textwidth]{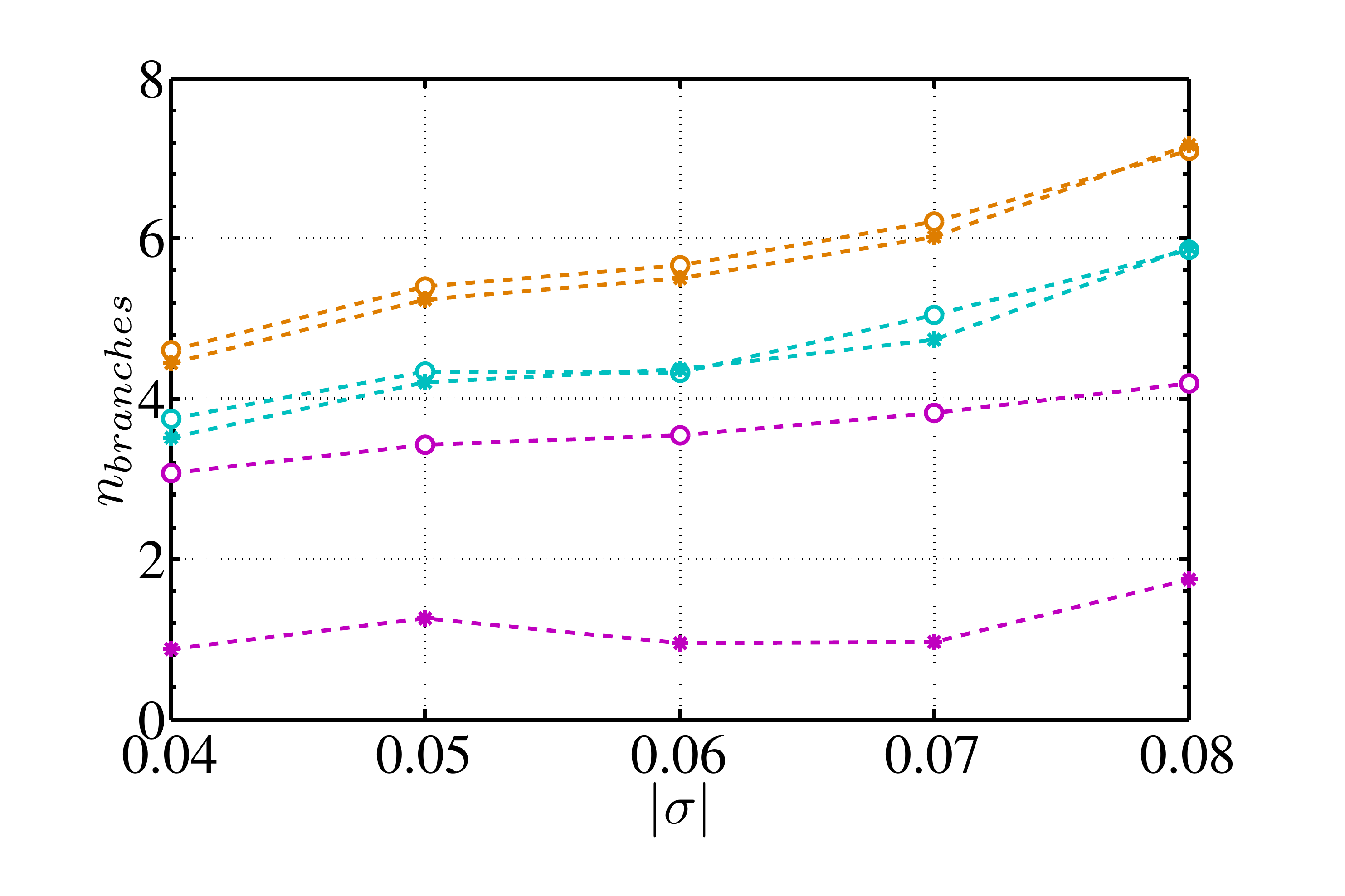}}
\caption{(color online) Average number of branches of the plectonemes as a
function of the absolute value of $\sigma$, for chains
with high/low rigidities compared to chains with a single rigidity given by the weighted average of the matching pair of persistence lengths: $\langle
l_p \rangle = \frac{1}{3}l_p^{(l)} + \frac{2}{3}l_p^{(h)}$. The pair rigidities
and the corresponding weighted averges are: 
$(l_p^{(l)}, l_p^{(h)}) = (40nm, 55nm)$ (orange dashed line with circles) and
$50nm$ (orange solid line with asterisks);
$(55nm, 75nm)$ (blue dashed line with circles) and
$68.3nm$ (blue solid line with asterisks); 
$(50nm, 200nm)$ (violet dashed line with circles) and
$150nm$ (violet solid line with asterisks).}  
\label{fig:FigS3}
\end{figure}

\begin{figure}[tb]
\center{\includegraphics[width=0.5\textwidth]{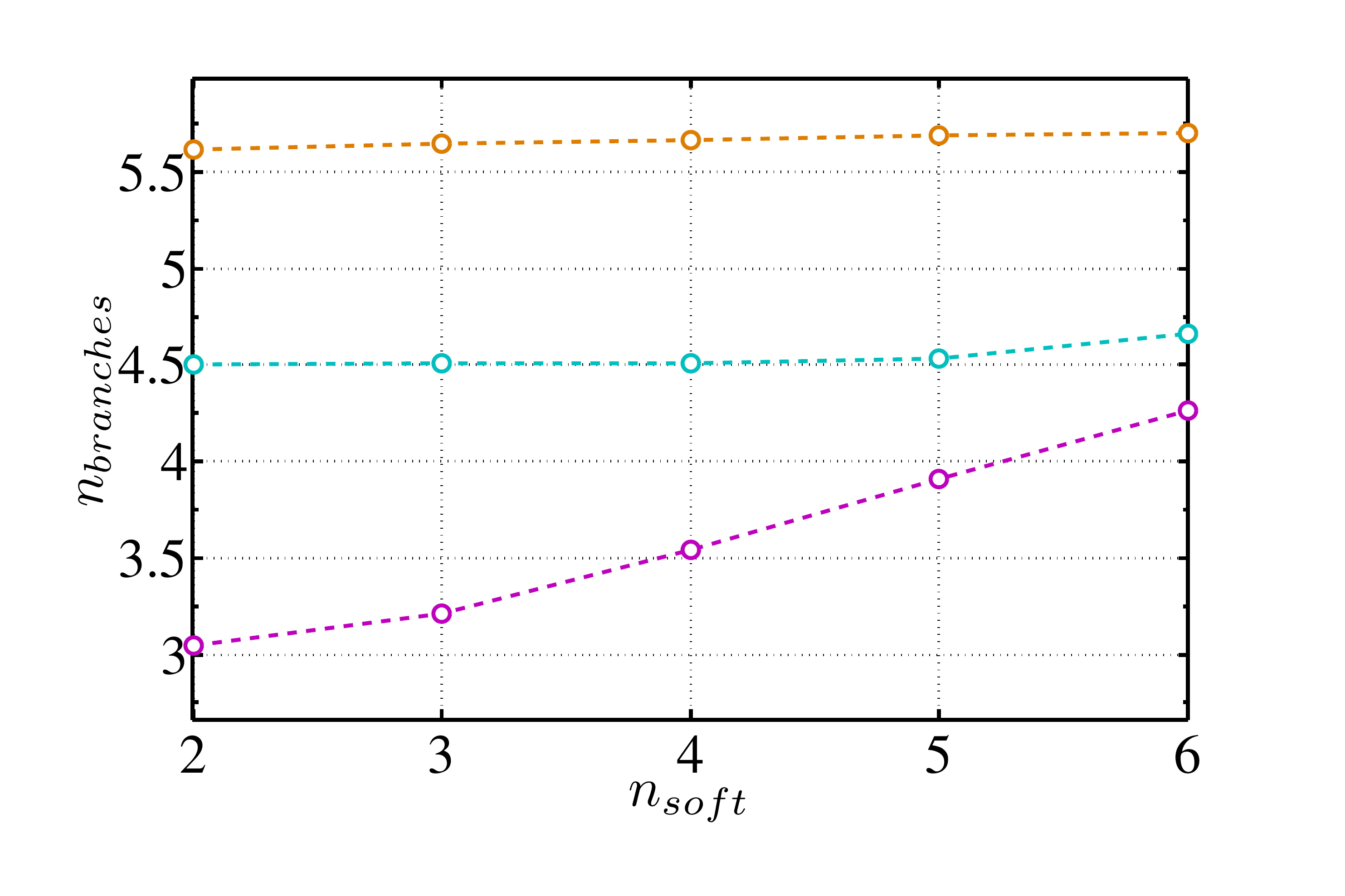}}
\caption{(color online) Average number of plectoneme branches as a
function of the number of soft domains for chains
with persistence-length pairs: $(l_p^{(l)}, l_p^{(h)}) = (40nm, 55nm)$
in orange, $(55nm, 75nm)$ in blue, and
$(50nm, 200nm)$ in violet.} 
\label{fig:FigS4}
\end{figure}

\begin{figure}[tbh]
\centering
\subfigure[] {\label{fig:figS5histA}
\includegraphics[scale=0.29]{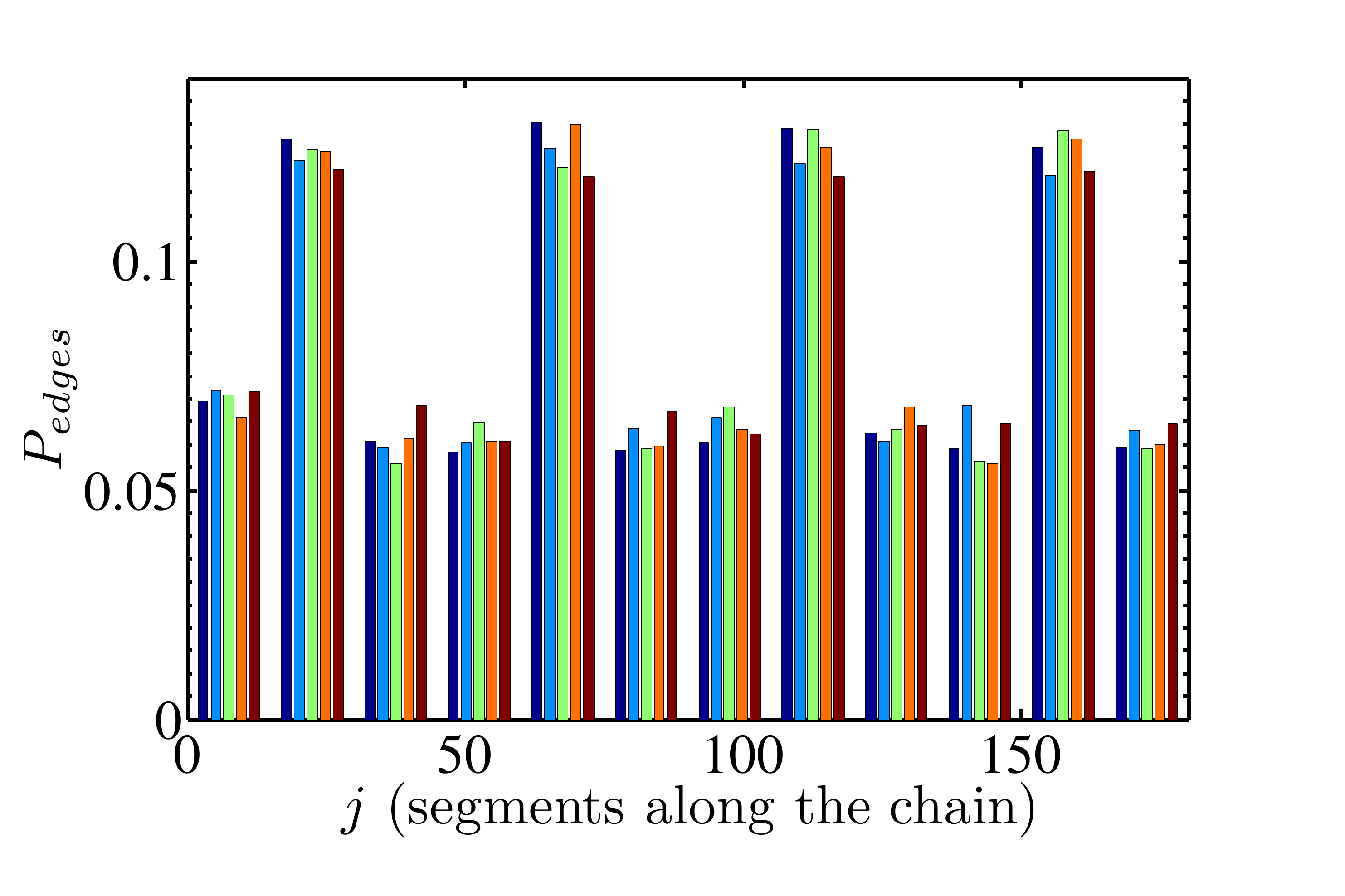}}
%\newline%
\centering
\subfigure[]{\label{fig:figS5histB}
\includegraphics[scale=0.29]{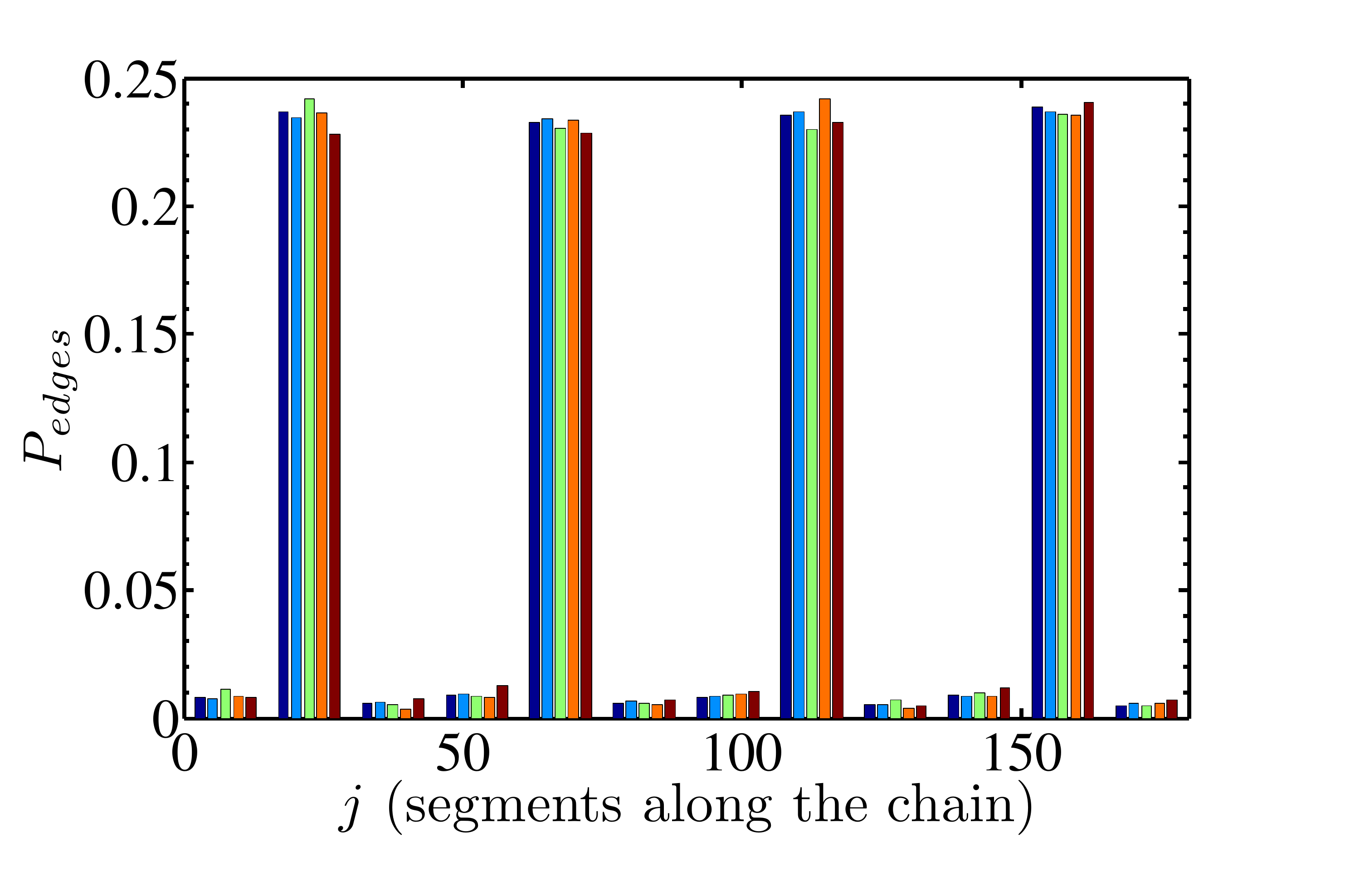}}
\caption{(Color online) Histograms of the locations of the centers of
the plectonemic edges (loops), $P_{edges}$, for a circular chain
with $4$ soft regions located around the $23$rd, $68$th,
$113$rd and $158$th segments for: $\sigma=-0.04$ (dark blue),
$\sigma=-0.05$ (light blue), $\sigma=-0.06$ (light green), $\sigma=-0.07$
(orange) and $\sigma=-0.08$ (red). (a) $(l_p^{(l)}, l_p^{(h)}) = (40nm,
55nm)$, (b) $(l_p^{(l)}, l_p^{(h)}) = (50nm, 200nm)$. Note
that the probability of localization of the edges to the kinks doesn't change significantly with
$\sigma$.}
\label{fig:figS5} 
\end{figure}

\begin{figure}[tbh]
\centering
\subfigure[] { \label{fig:figS6hist2A}
\includegraphics[scale=0.29]{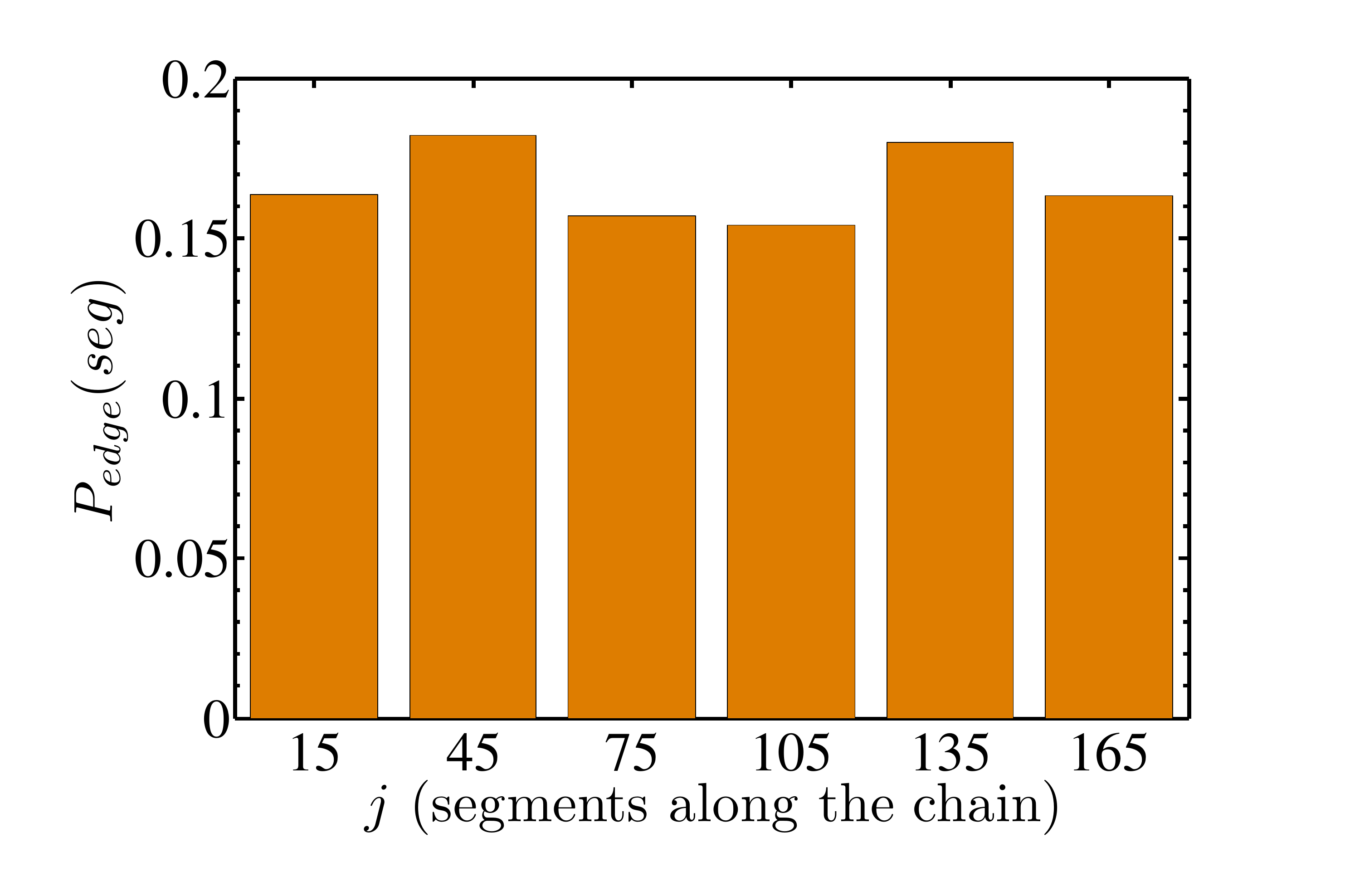}}
%\newline%
\centering
\subfigure[]{\label{fig:figS5hist3B}
\includegraphics[scale=0.29]{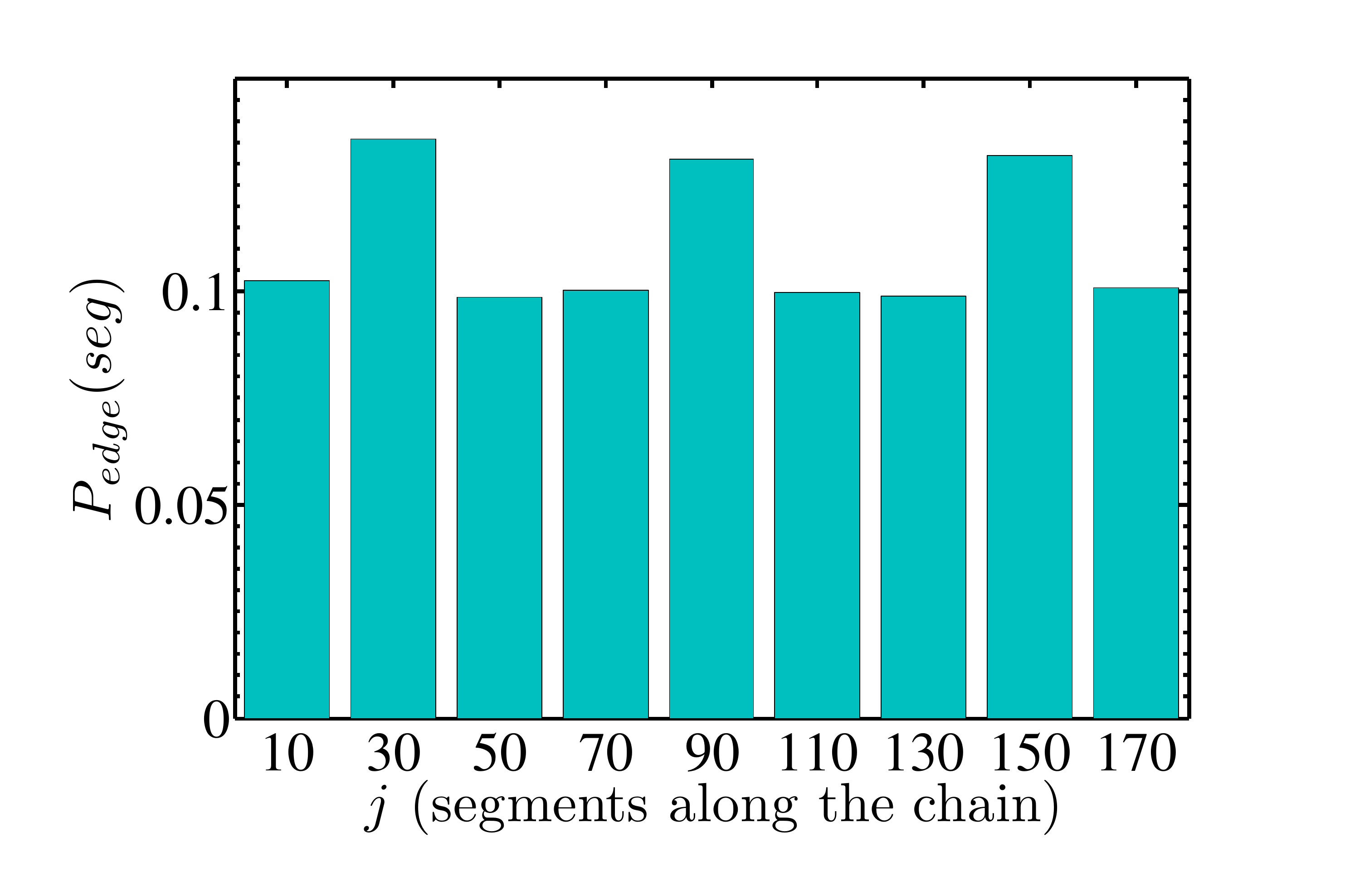}}
\newline%
\centering
\subfigure[]{\label{fig:figS5hist4B}
\includegraphics[scale=0.29]{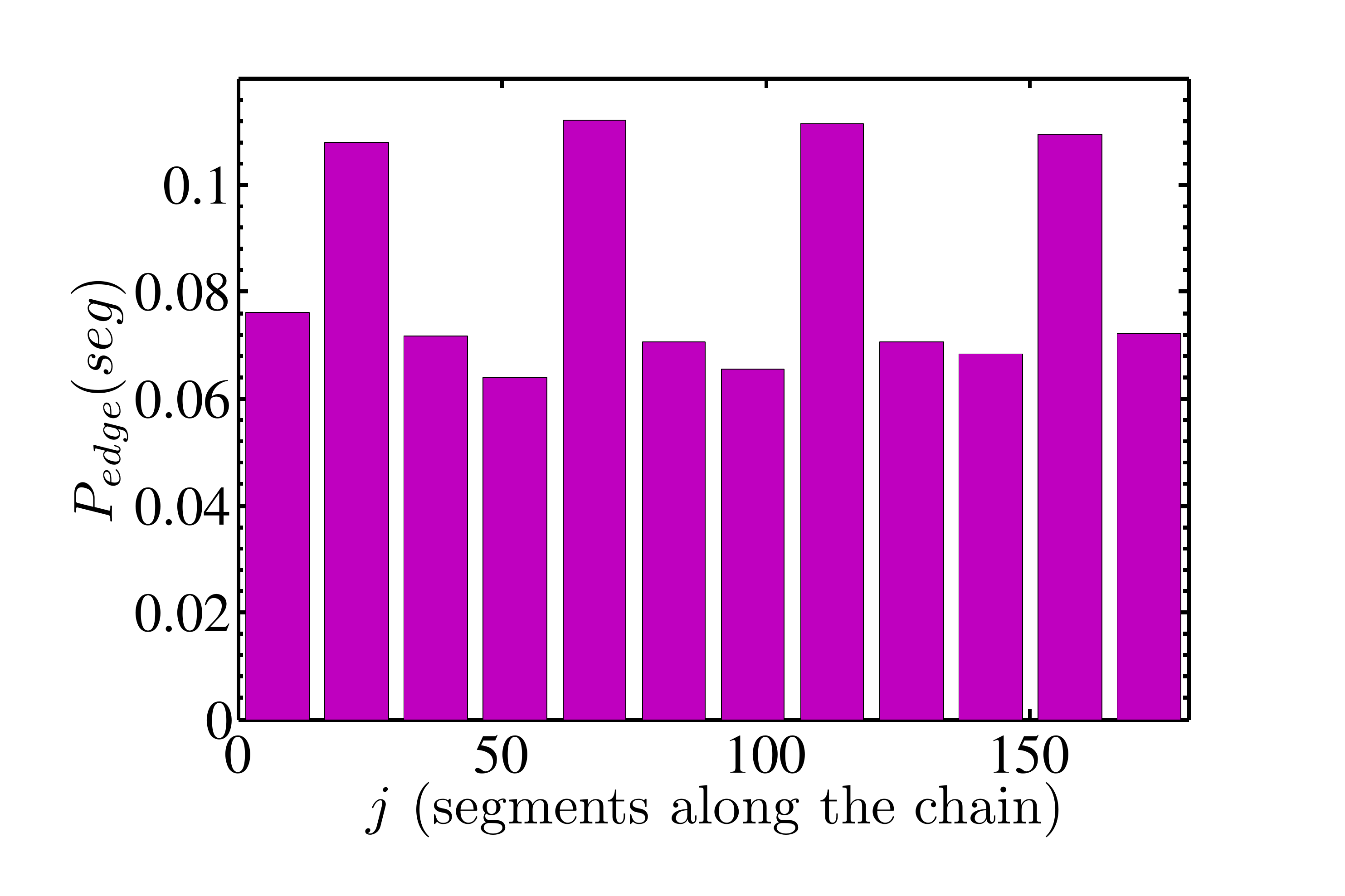}}
%\newline%
\centering
\subfigure[] { \label{fig:figS5hist5A}
\includegraphics[scale=0.29]{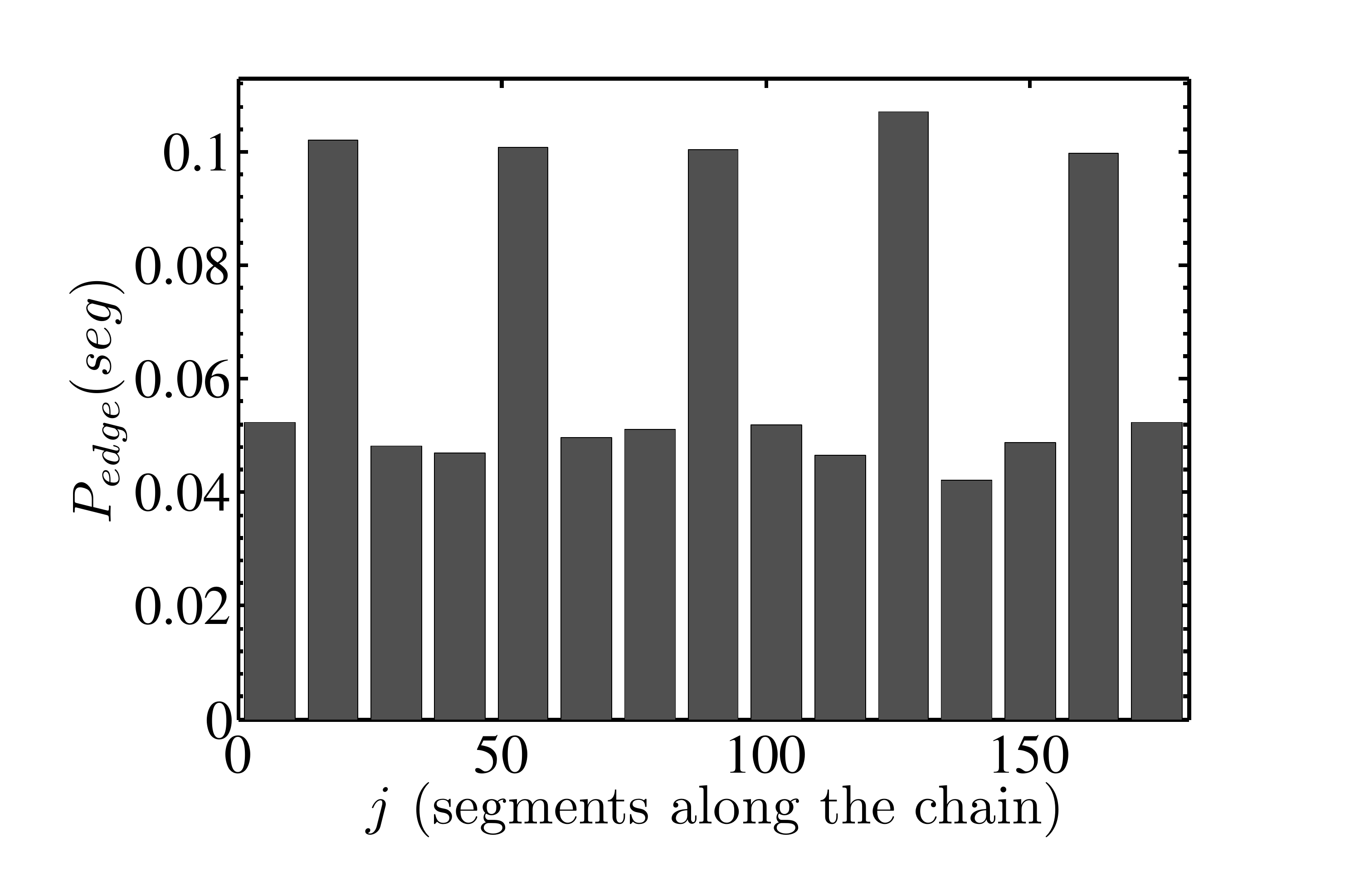}}
\newline%
\centering
\subfigure[]{\label{fig:figS5hist6B}
\includegraphics[scale=0.29]{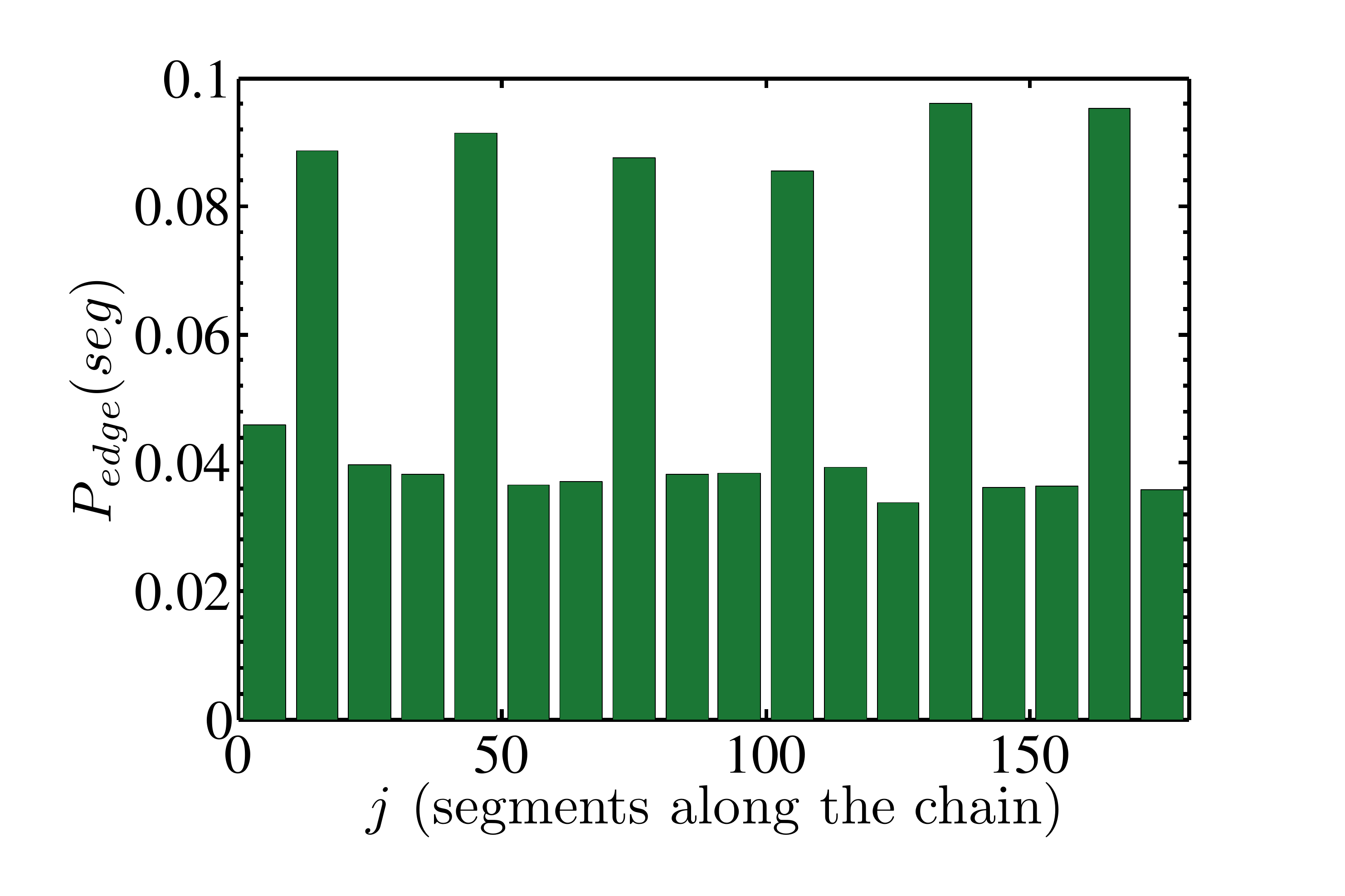}}
%\newline%
\centering
\subfigure[]{\label{fig:figS6histTot}
\includegraphics[scale=0.29]{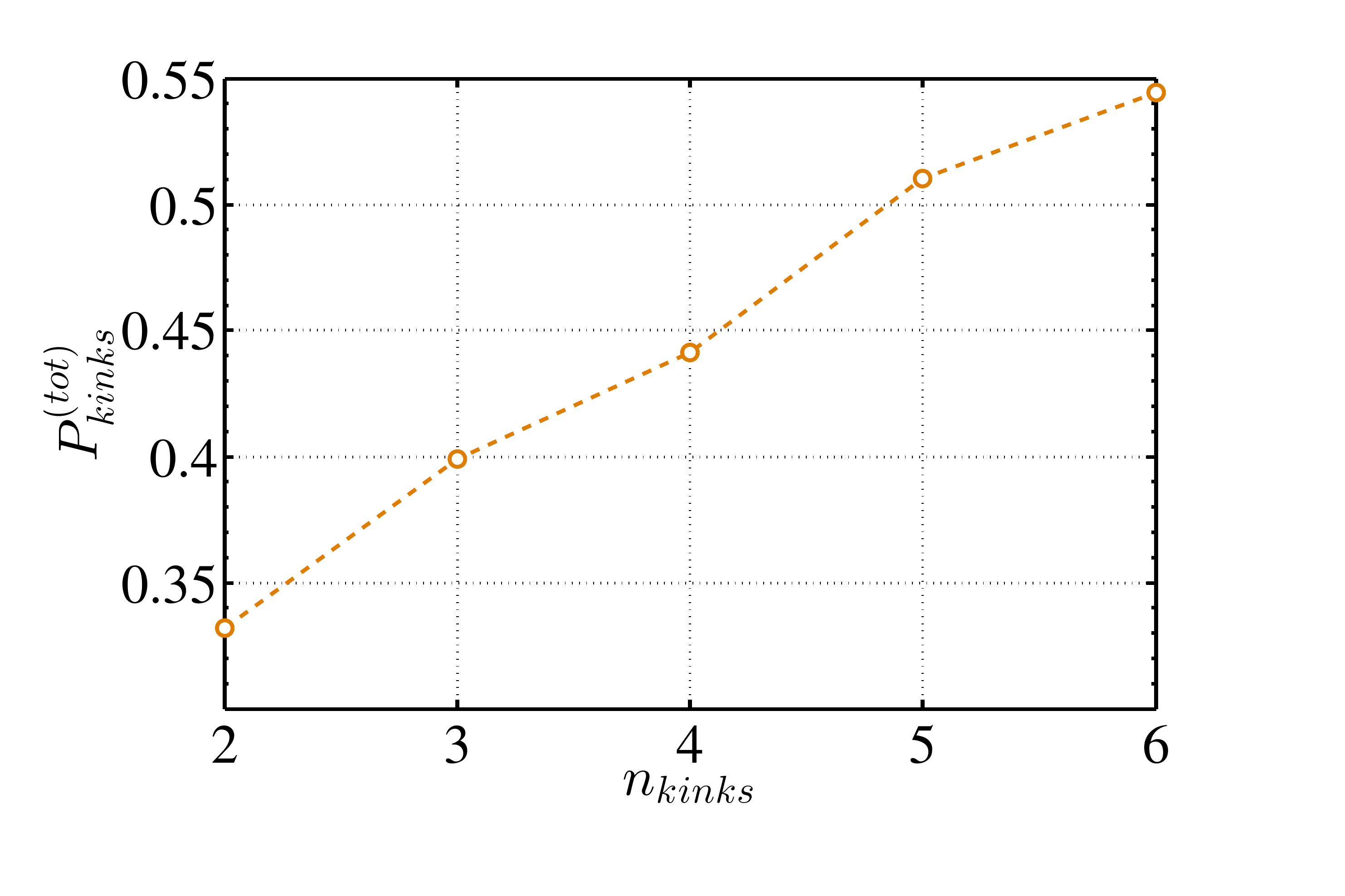}}
\caption{(Color online) Histograms of the locations of the centers of
the plectoneme edges, $P_{edges}$, for a circular chain of
$L\simeq5400bp\simeq180seg$
with point-like kinks with $l_p=4nm$ where the rest of the chain has
$l_p=50nm$. (a) two kinks located at $j=45$ and $j=135$ (b) three kinks
located at $j=31$, $j=91$ and $j=151$ (c) four kinks
located at $j=23$, $j=68$, $j=113$ and $j=158$ (d) five kinks
located at $j=19$, $j=55$, $j=91$, $j=127$ and $j=163$ (e) six kinks
located at $j=16$, $j=46$, $j=76$, $j=106$, $j=136$ and $j=166$ (f) the overall
probability for an edge to contain a kink as a function of the number
of kinks. The existence of higher number of kinks effectively ``sucks'' the
probability from the other parts of the chain, and as can be seen in (a)-(e) the
difference between the probabilities of kink-containing and kink-free edges increases.
}
\label{fig:figS6}
\end{figure}

\begin{figure}[tb]
\center{\includegraphics[width=0.5\textwidth]{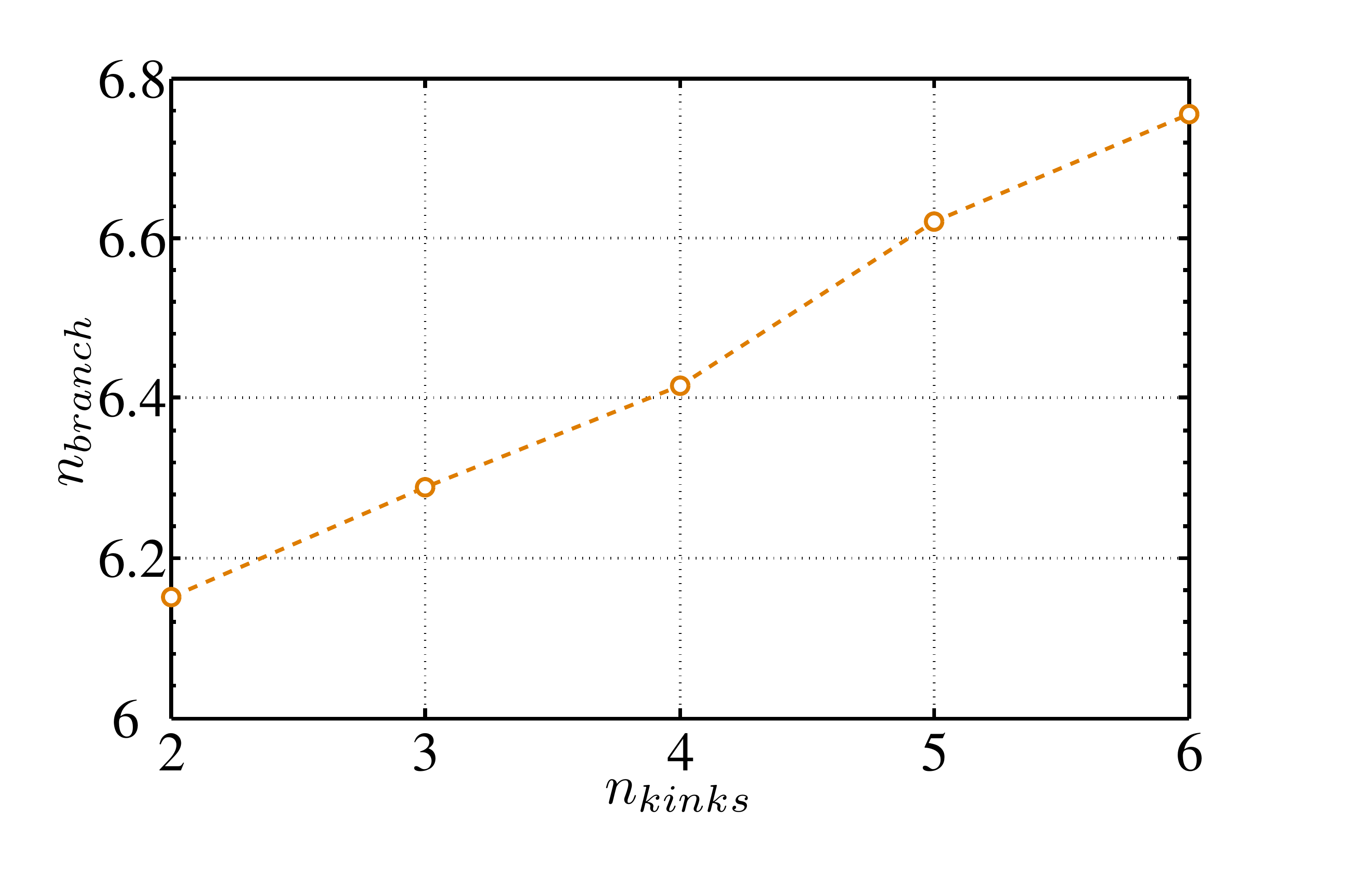}}
\caption{(color online) The mean number of plectoneme branches as a function of
the number of kinks for $5400bp$-long chains with $l_p=50nm$ for the chain and
$l_p=4nm$ for the kinks.} 
\label{fig:FigS7}
\end{figure}

\begin{figure}[tb]
\center{\includegraphics[width=0.5\textwidth]{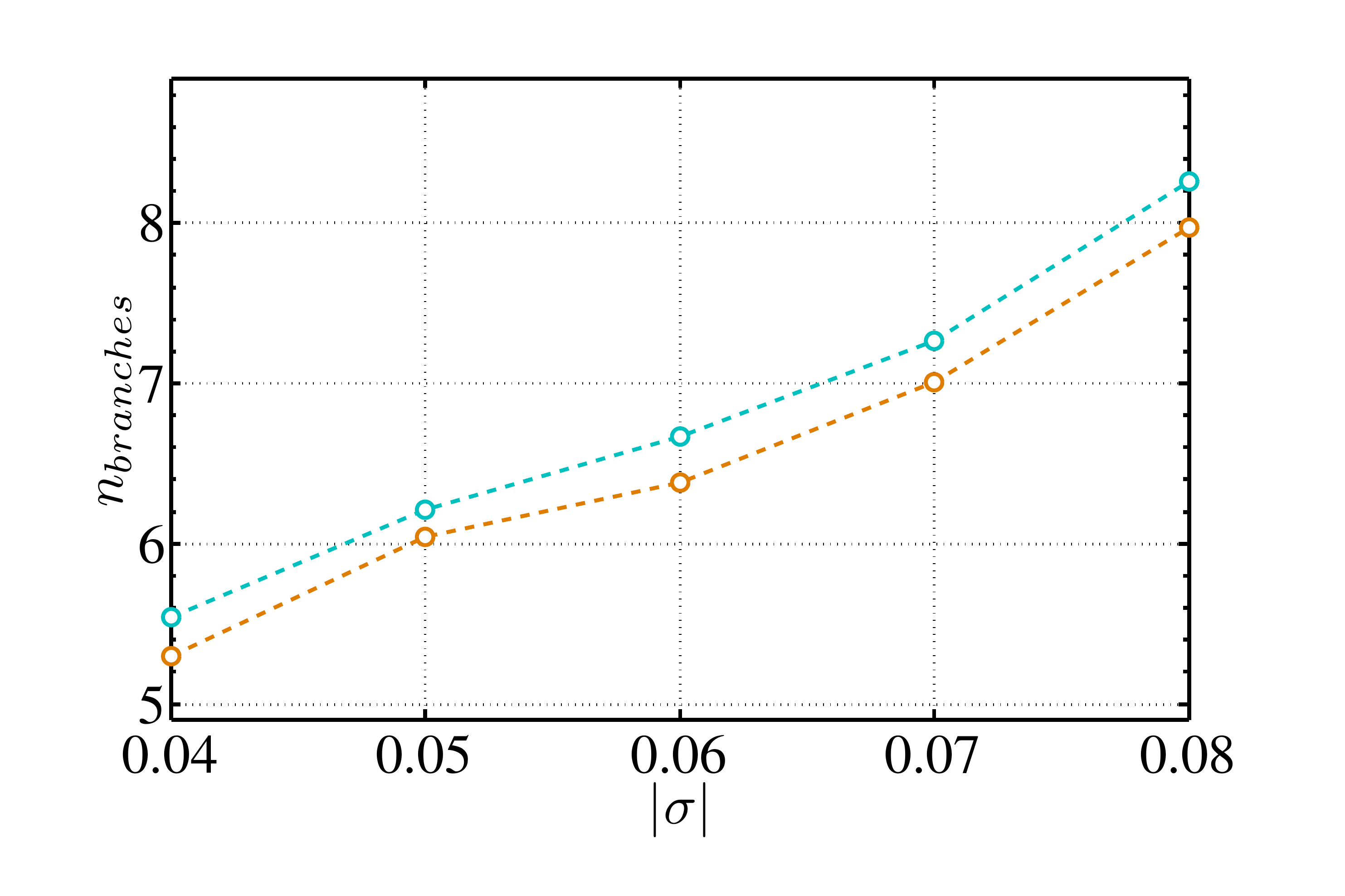}}
\caption{(color online) The number of branches of the chain, $n_{branch}$, as a
function of $|\sigma|$ for $4$-kinks (orange) and $6$-kinks (light blue). While
for $6$-kinks the number of branches is comparable with the number of kinks,
for $4$-kinks the number of branches exceeds the number of kinks.} 
\label{fig:FigS8}
\end{figure}

\begin{figure}[tbh]
\centering
\subfigure[] { \label{fig:figS8hist4}
\includegraphics[scale=0.29]{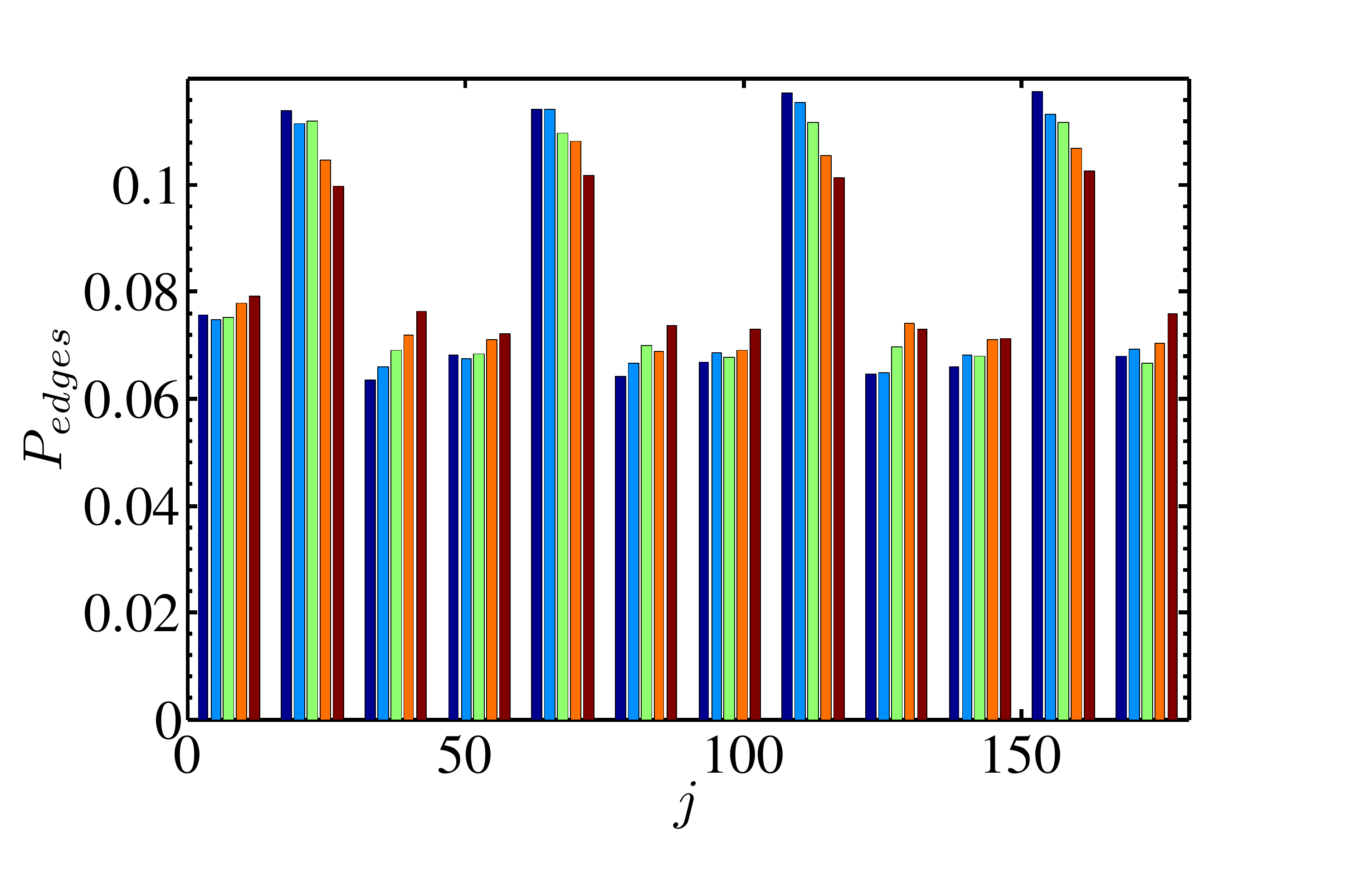}}
%\newline%
\centering
\subfigure[]{\label{fig:figS8hist6}
\includegraphics[scale=0.29]{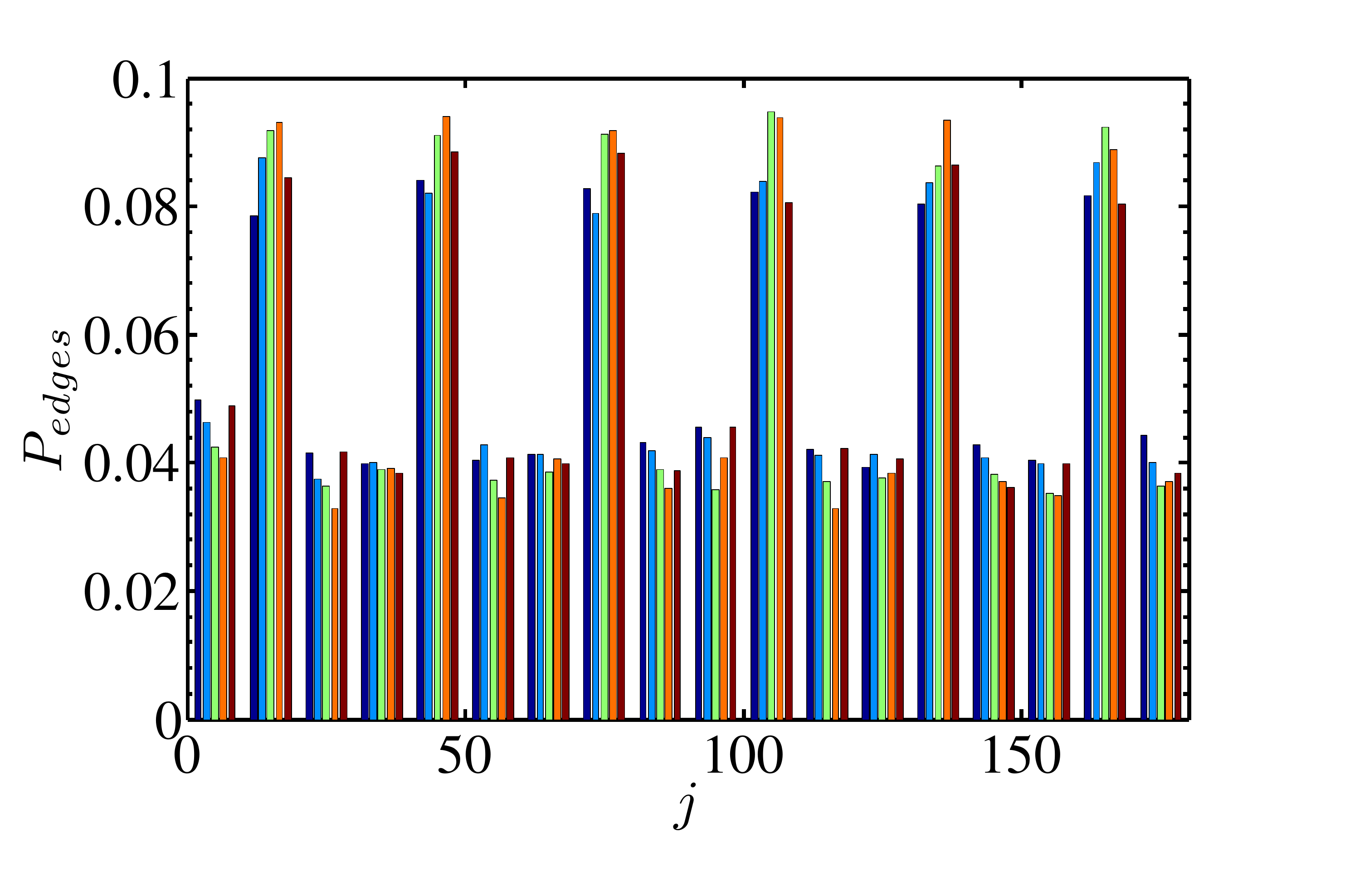}}
\caption{(Color online) Histograms of the locations of the plectoneme edges,
$P_{edges}$, for $\sigma=-0.04$ (dark blue), $\sigma=-0.05$
(light blue), $\sigma=-0.06$ (light green), $\sigma=-0.07$ (orange) and
$\sigma=-0.08$ (red). (a) $4$ kinks located at the $22$nd,
$62$nd, $112$nd and $152nd$ dimers and (b) $6$ kinks
located at $j=16$, $j=46$, $j=76$, $j=106$, $j=136$ and $j=166$.}
\label{fig:FigS9}
\end{figure}

\begin{figure}[tbh]
\centering
\subfigure[] { \label{fig:figS10distHistA}
\includegraphics[scale=0.31]{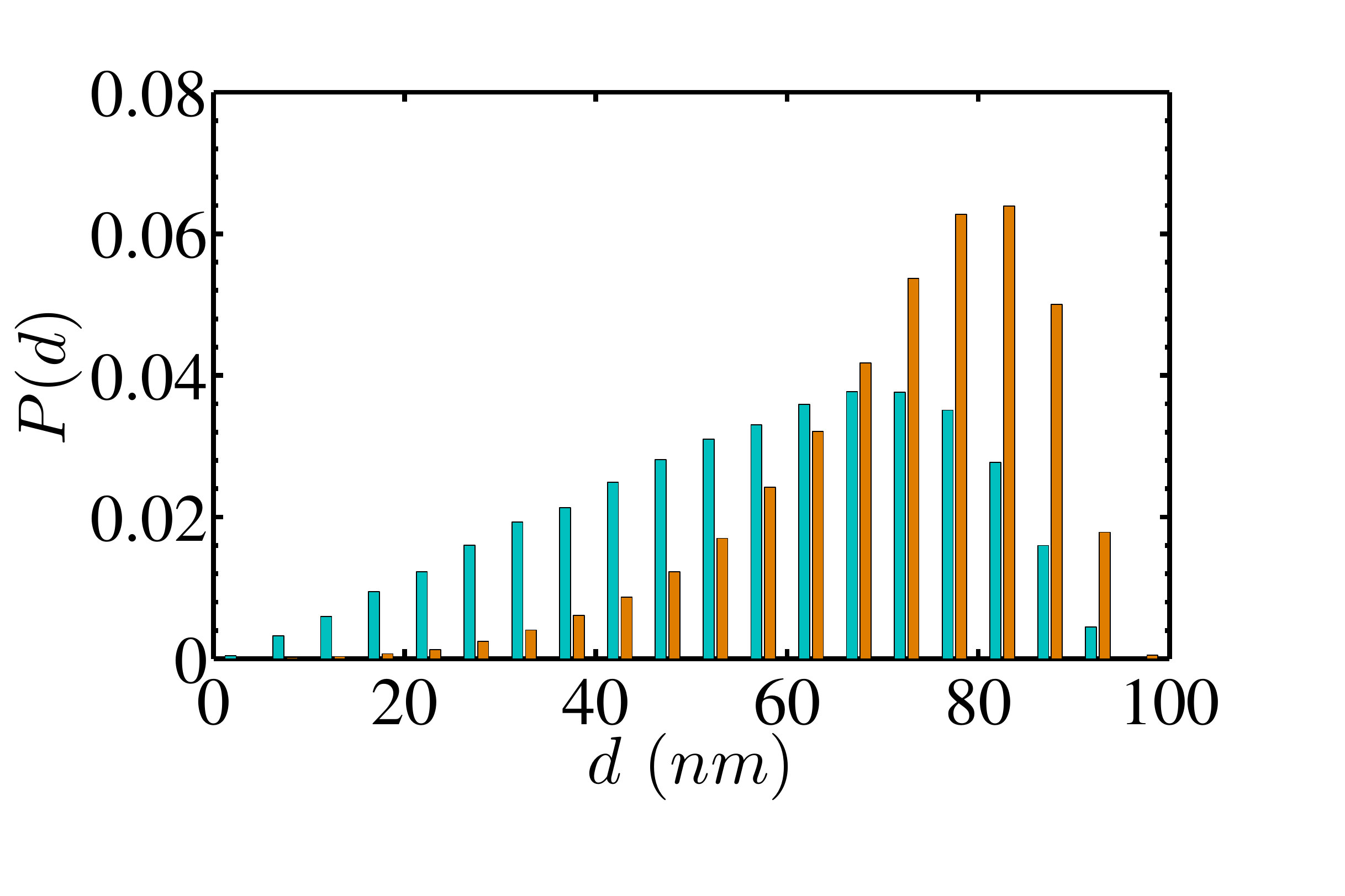}}
%\newline%
\centering
\subfigure[]{\label{fig:figS10distHistB}
\includegraphics[scale=0.31]{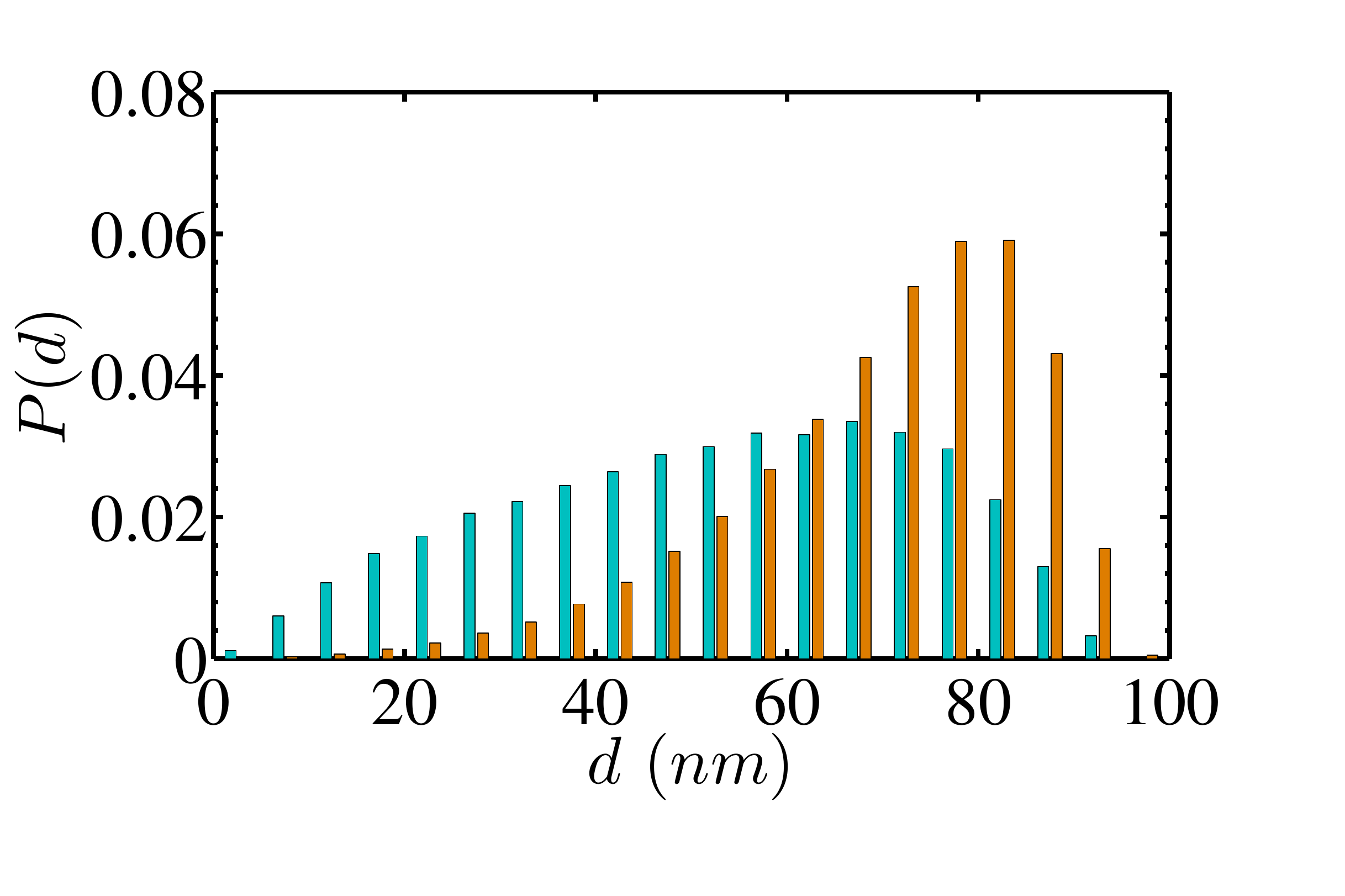}}
\newline%
\centering
\subfigure[]{\label{fig:figS10distHistC}
\includegraphics[scale=0.31]{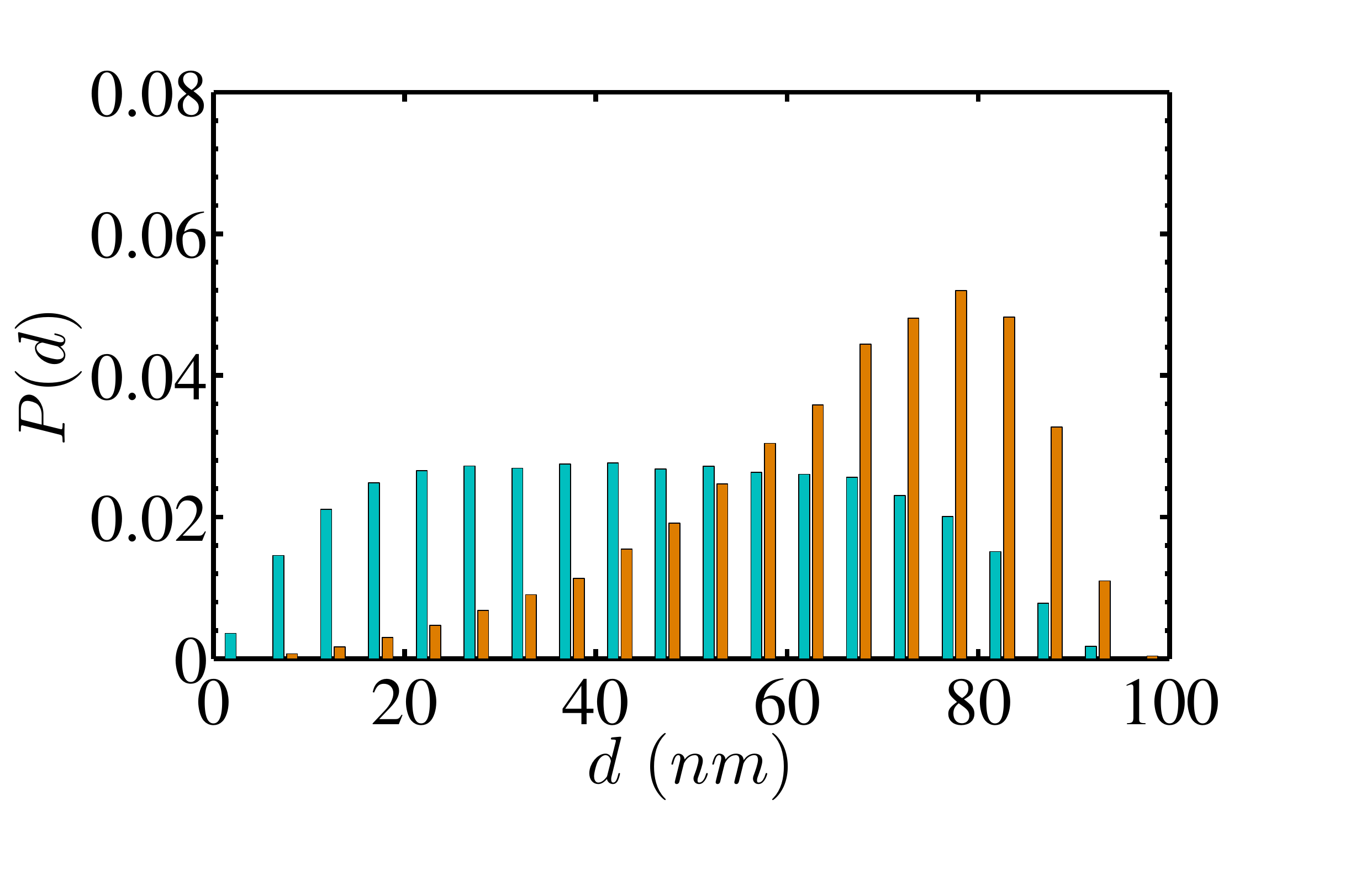}}
%\newline%
\centering
\subfigure[] { \label{fig:figS10distHistD}
\includegraphics[scale=0.31]{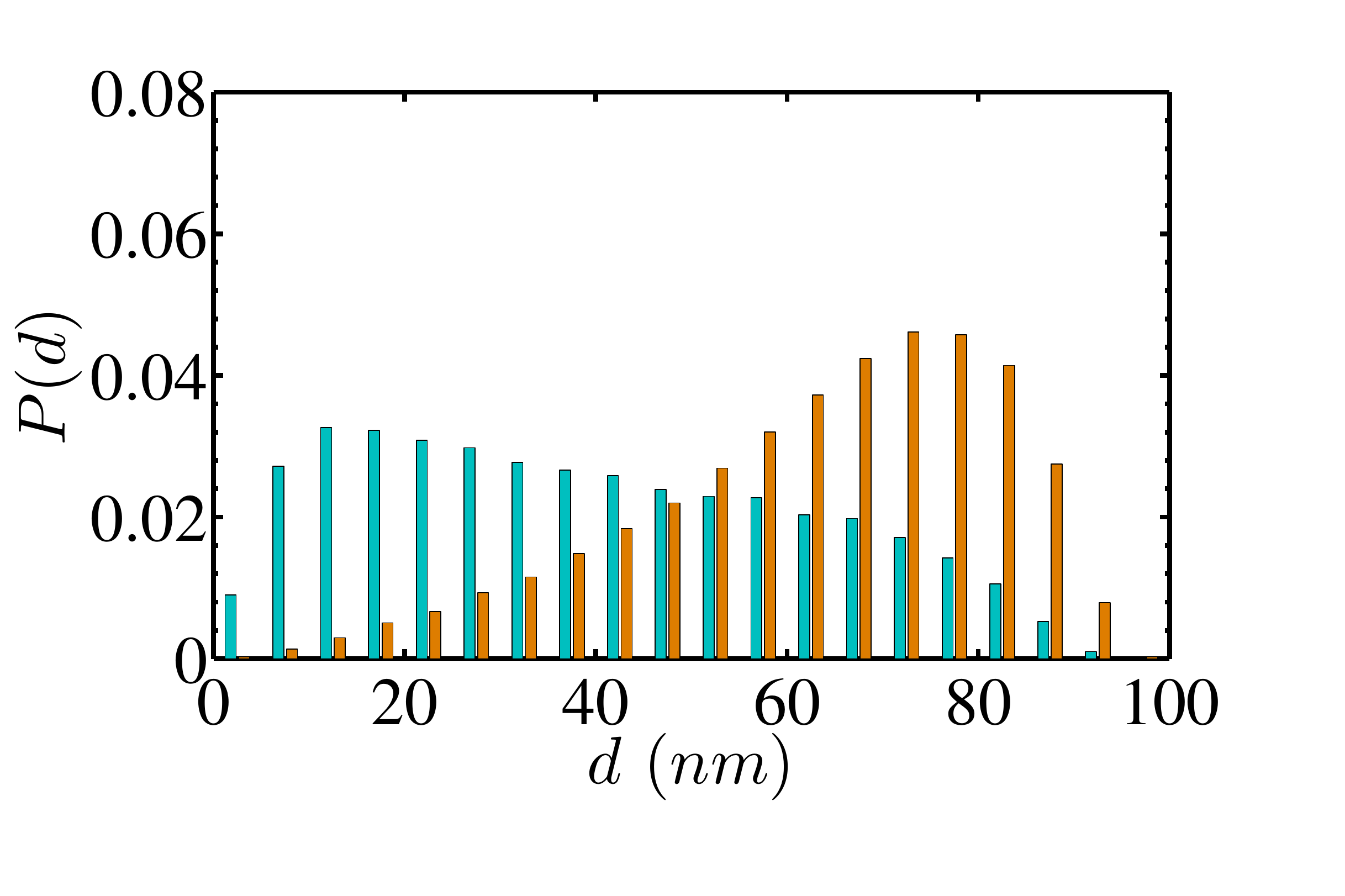}}
\newline%
\centering
\subfigure[]{\label{fig:figS10distHistE}
\includegraphics[scale=0.31]{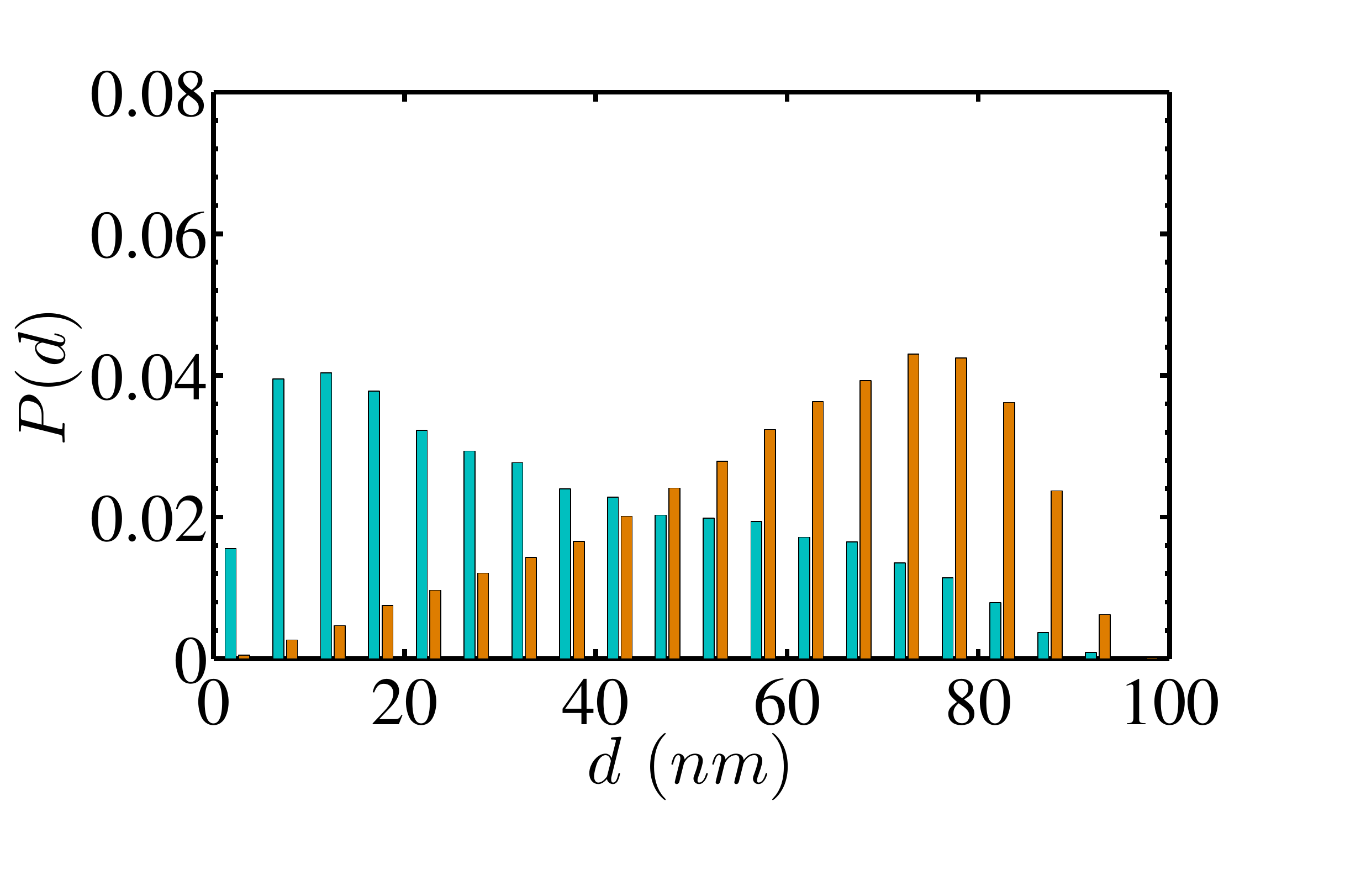}}
%\newline%
\centering
\subfigure[]{\label{fig:figS10distHistF}
\includegraphics[scale=0.31]{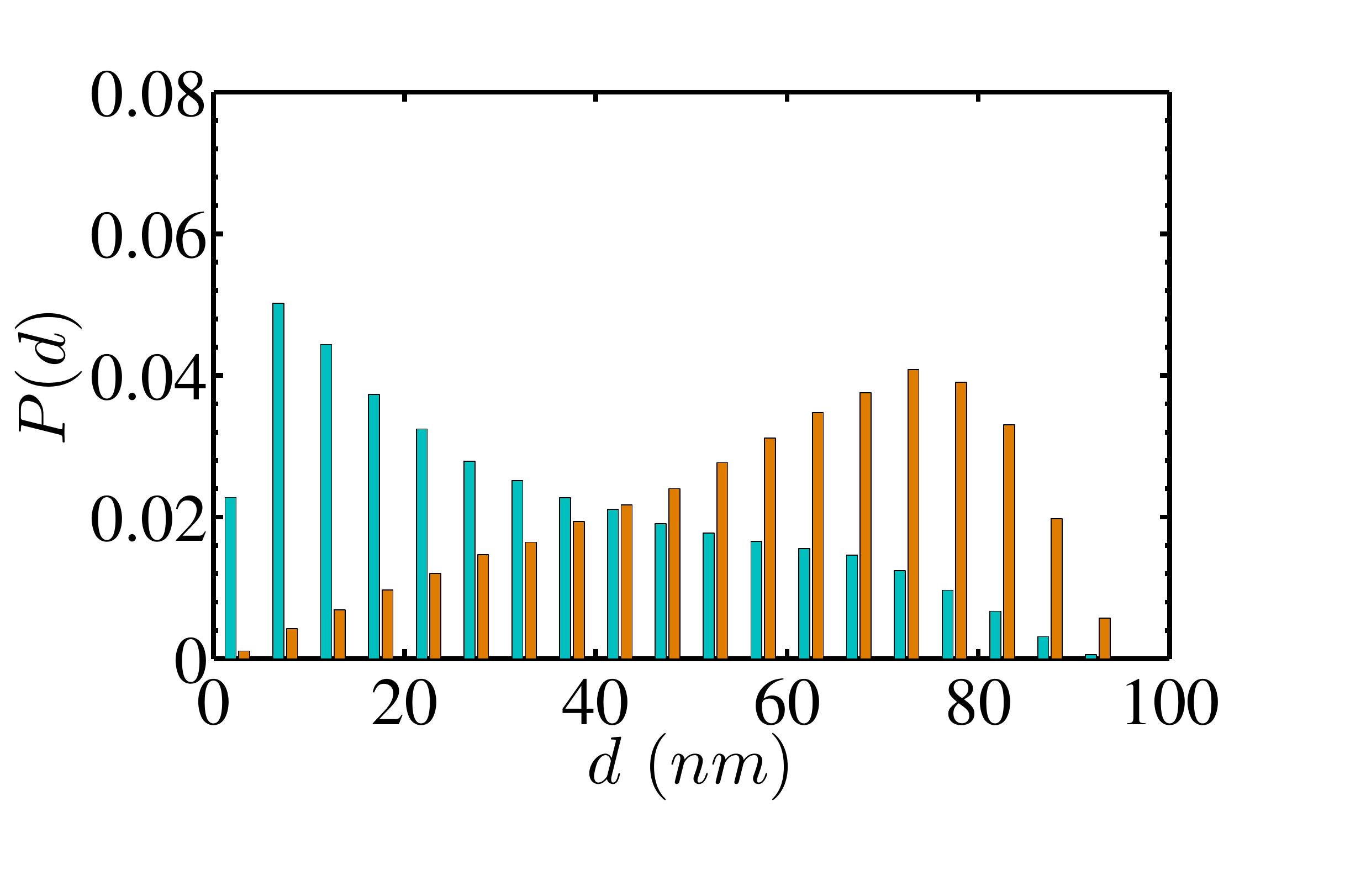}}
\newline%
\centering
\subfigure[]{\label{fig:figS10distHistG}
\includegraphics[scale=0.31]{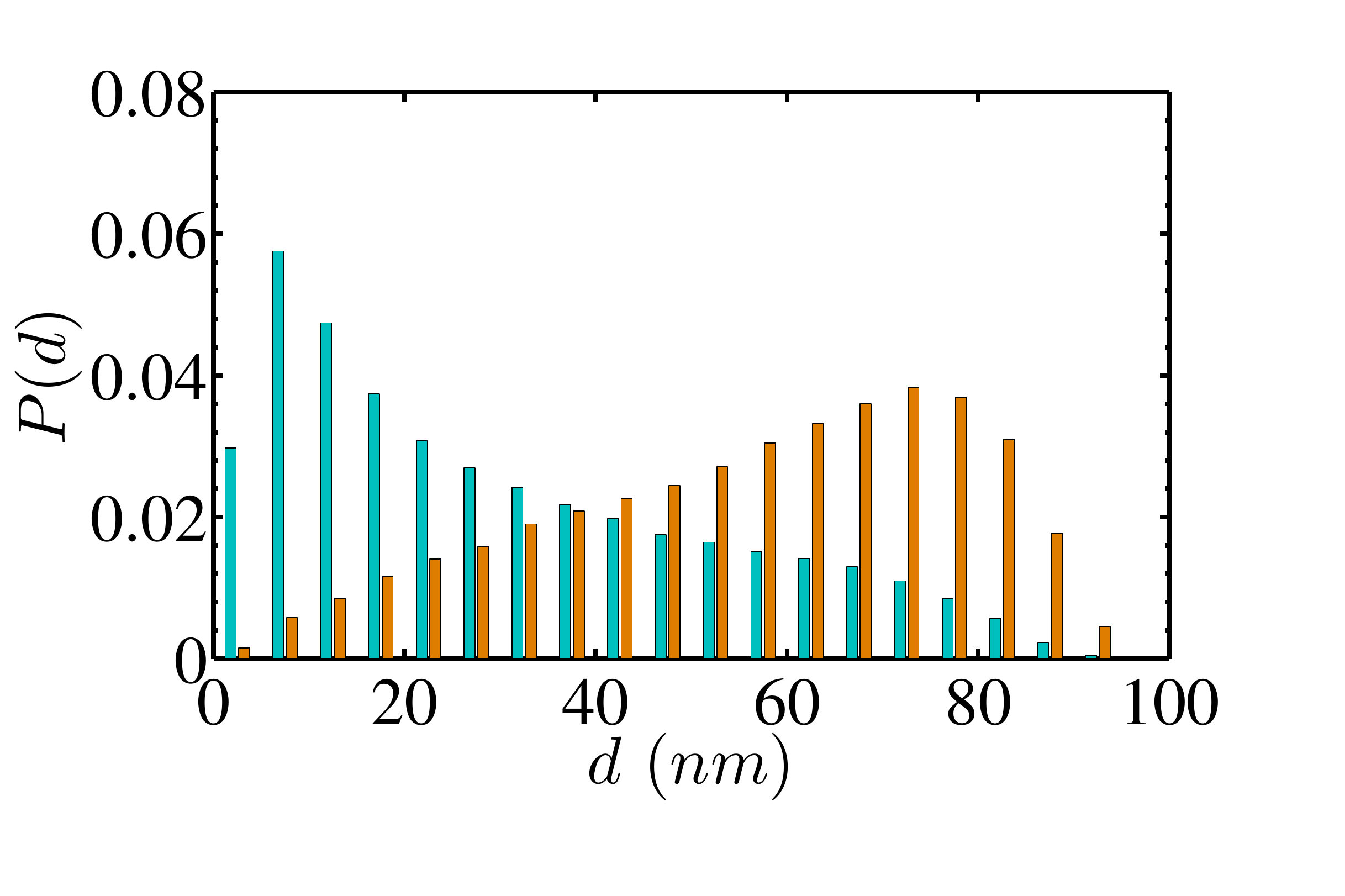}}
%\newline%
\centering
\subfigure[] { \label{fig:figS10distHistH}
\includegraphics[scale=0.31]{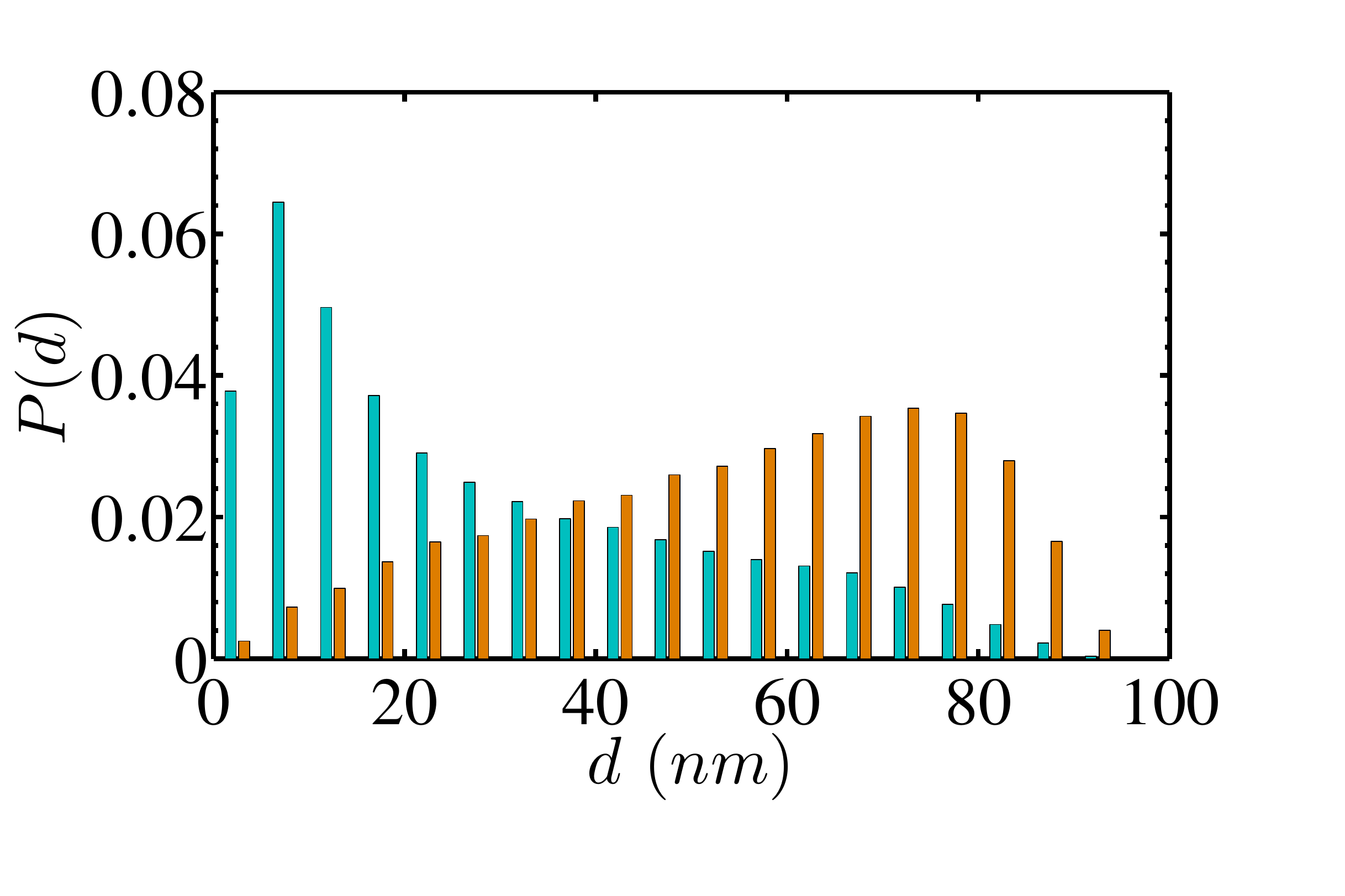}}
\caption{(Color online) Histograms of distances of the end points of a
$100nm$ long segment in a $5400bp$ plasmid containing a kink (light blue) and
not containing a kink (orange), for (a) $\sigma=0$ (b) $\sigma=-0.01$ (c)
$\sigma=-0.02$ (d) $\sigma=-0.03$ (e) $\sigma=-0.04$ (f) $\sigma=-0.05$ (g)
$\sigma=-0.06$ (h) $\sigma=-0.07$. The shift of the maximal probability to
short distances, and the significant enhancement of probability for the
shortest distances are clearly observed for segments containing a kink.}
\label{fig:figS2}
\end{figure}

\end{document}